%% file: main.tex
\documentclass[sigconf, nonacm]{acmart}

\input{macros}

\usepackage{makecell}
\usepackage[autostyle, english = american]{csquotes}
\MakeOuterQuote{"}
\usepackage{subcaption}
\usepackage{listings}
\usepackage{xcolor}
\usepackage{smartdiagram}
\usepackage{tikz}
\usepackage{pgf-pie}
\usepackage{cleveref}
\usepackage{booktabs}
\usepackage{tabularx}
\usepackage{pifont}
\newcommand{\cmark}{\ding{51}}%
\newcommand{\xmark}{\ding{55}}%
\usepackage{optidef}
\usepackage{algorithm}
\usepackage{algpseudocode}
\usepackage{xspace}
\usepackage{float}
\usepackage{pgf}
\usepackage{pgfplots}
 
  \everymath=\expandafter{\the\everymath\displaystyle}
  \usepackage{amsmath}
  \makeatletter\@ifpackageloaded{underscore}{}{\usepackage[strings]{underscore}}\makeatother
\usepackage{amsmath}
\usepackage{url}
\usepackage{multirow}
\newcommand{\notimplies}{%
  \mathrel{{\ooalign{\hidewidth$\not\phantom{=}$\hidewidth\cr$\implies$}}}}
\usetikzlibrary{positioning, shapes.geometric, arrows}
\usetikzlibrary{shapes, arrows.meta, shapes.geometric, fit, backgrounds}
\definecolor{lightblue}{rgb}{0.68, 0.85, 0.9}
\definecolor{codegreen}{rgb}{0,0.6,0} 
\definecolor{darkblue}{rgb}{0.0, 0.0, 0.5}
\definecolor{darkblueilp}{RGB}{0, 0, 175}
\definecolor{darkred}{rgb}{0.5, 0.0, 0.0}
\definecolor{darkgreen}{rgb}{0.0, 0.5, 0.0}
\definecolor{lightgray}{gray}{0.95}
\usepackage{tcolorbox}
 \usepackage{balance}
\usepackage{threeparttable} 

\tcbuselibrary{listings, skins}
\newtcblisting{mypython}{
  left=1mm,    
  right=1mm,   
  top=0mm,     
  bottom=0mm,  
  listing only,
  listing options={
    language=Python,
    basicstyle=\ttfamily\footnotesize,
    keywordstyle=\color{darkblue},
    stringstyle=\color{darkred},
    commentstyle=\color{darkgreen},
    breaklines=true,
    breakindent=0pt,
    postbreak=\mbox{\textcolor{darkred}{\footnotesize$\hookrightarrow$}\space},
    showstringspaces=false,
    lineskip=-1pt
  },
  boxrule=0.5pt,
  colback=lightgray, 
  colframe=black,    
}
\usepackage{mathtools}

\newcommand\vldbdoi{XX.XX/XXX.XX}

\newcommand\vldbavailabilityurl{URL_TO_YOUR_ARTIFACTS}
\newcommand\vldbpagestyle{plain}

\begin{document}

\title{\spade: Synthesizing Data Quality Assertions \\for Large Language Model Pipelines~\papertext{[Industry]}}


\author{Shreya Shankar$^1$, Haotian Li$^2$, Parth Asawa$^1$, Madelon Hulsebos$^1$, Yiming Lin$^1$, J.D. Zamfirescu-Pereira$^1$, Harrison Chase$^3$, Will Fu-Hinthorn$^3$, Aditya G. Parameswaran$^1$, Eugene Wu$^4$}
\affiliation{%
$^1$UC Berkeley, $^2$HKUST, $^3$LangChain, $^4$Columbia University \\
\{\url{shreyashankar, pgasawa, madelon, yiminglin, zamfi, adityagp}\} \url{@ berkeley.edu} \\
\url{haotian.li @ connect.ust.hk}, \{\url{harrison, wfh}\} \url{@ langchain.dev},  \url{ewu @ cs.columbia.edu}}




\begin{abstract}
Large language models (LLMs) are
being increasingly deployed as part of pipelines
that repeatedly process or generate data of some sort.
However, a common barrier to deployment
are the frequent and often unpredictable errors
that plague LLMs.
Acknowledging the inevitability of these
errors, we propose {\em data quality assertions}
to identify when LLMs may be making mistakes.  
We present \spade, a method for automatically synthesizing 
data quality assertions that identify bad LLM outputs. 
We make the observation that developers 
often identify data quality issues 
during prototyping prior
to deployment,
and attempt to address them by adding instructions
to the LLM prompt over time. 
\spade therefore 
analyzes histories of prompt versions over time 
to create candidate assertion functions and then selects a minimal set that fulfills both coverage and accuracy requirements. 
In testing across nine different real-world LLM pipelines, \spade efficiently reduces the number of assertions by 14\% and decreases false failures by 21\% when compared to simpler baselines.
\spade has been deployed as an offering within LangSmith,
LangChain's LLM pipeline hub, and has been used to generate
data quality assertions for over 2000
pipelines across a spectrum of industries. 
\end{abstract}

\maketitle

\pagestyle{\vldbpagestyle}


\section{Introduction}

\input{sections/intro}


\section{Identifying Candidate Assertions}

\input{sections/candidateassertions}

\section{Filtering Candidate Assertions}

\input{sections/methods}

\section{Evaluation}

\input{sections/experiments}

\section{Related Work}

\input{sections/related}

\section{Conclusion and Future Work}

We introduce a new problem of auto-generating assertions
to catch failures in LLM pipelines,
as well as \spade, our framework
for doing so. \spade
comprises two components: first, it synthesizes
candidate assertions, and then it filters
them down into a more manageable subset.
To synthesize candidate assertions, we analyzed prompt version
histories and learned that prompt deltas
are often a rich source of requirements and therefore candidate
assertions. We developed
a taxonomy of prompt deltas for assertion synthesis,
demonstrating its value via integration and deployment
with LangChain, with over 2000 runs across domains.
For the latter problem of candidate
assertion filtering,
we expressed the selection of an optimal set of assertions,
covering most failures while introducing as few false failures
as possible as an Integer Linear Program (ILP).
We proposed assertion subsumption to cover failures in 
data-scarce scenarios and incorporated this into our ILP. 
We also studied the setting where there are no examples
and demonstrated that it reduces to a topological sort
on the subsumption graph.
Our auto-generating assertion system, \spade, 
was evaluated on nine real-world data-generating LLM pipelines. 
We have made our code and datasets public for further research and analysis.

There are a number of open questions in our effort
to make production LLM pipelines more robust.
For instance, while meeting developer provided criteria ($\alpha, \tau$)
is helpful,
sometimes developers would like to examine the generated and selected
assertions in a way that helps them make the tradeoffs themselves,
motivating a human-in-the-loop interface to {\em assist} developers in defining data quality for LLM pipelines. 
Such an interface could also be a vehicle for getting
developers to label examples on the fly. Determining
which labeled examples would help best select from the set of assertions
is an open question that is reminiscent of active learning.
Finally, we could also envision automatically updating assertion sets in deployed pipelines, as new failure modes inevitably arise in production.

\balance
\bibliographystyle{ACM-Reference-Format}
\bibliography{sample-base}

\clearpage

\techreport{\input{sections/app}}

\end{document}

%% file: macros.tex
\newcommand{\topic}[1]{\vspace{-3.5pt}\smallskip \smallskip \noindent{\bf #1.}}
\newcommand{\subtopic}[1]{\vspace{-3.5pt}\smallskip \smallskip \noindent{\em #1.}}

\newcommand{\techreport}[1]{#1}
\newcommand{\papertext}[1]{}

\theoremstyle{definition}

\newcommand*\circled[1]{\tikz[baseline=(char.base)]{
            \node[shape=circle,draw,inner sep=2pt] (char) {#1};}}

\newcommand{\spade}{\textsc{spade}\xspace}

\def\dashedrule#1#2#3{{%
\dimen1=#2 \divide\dimen1 by 2
\def\@ruledash{%
\rule{\dimen1}{0pt}%
\rule[0.5ex]{#1}{0.4pt}%
\rule{\dimen1}{0pt}}%
\count1=0
\loop%
\ifnum\count1<#3%
\advance\count1 by 1%
\@ruledash%
\repeat}}

%% file: sections/intro.tex

\label{sec:intro}

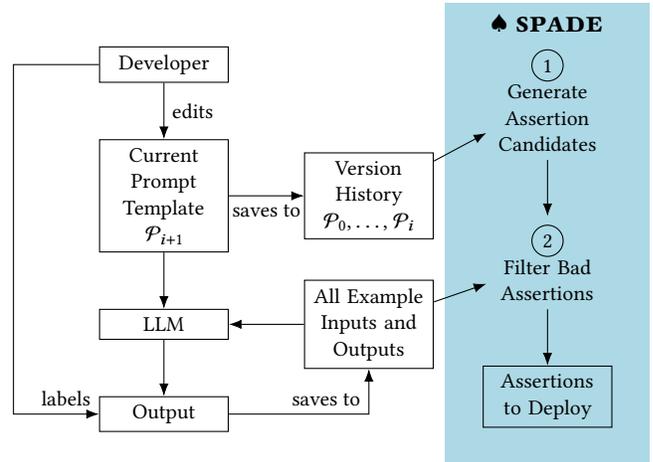
\begin{figure}[t]
\centering
\begin{tikzpicture}[
    node distance=0.75cm and 1cm,
    auto,
    block/.style={rectangle, draw, text width=1.5cm, align=center, font=\small},
    process/.style={text width=1.4cm, align=center, font=\small},
    line/.style={-Latex},
    spadeblock/.style={align=center, inner sep=2pt, font=\LARGE},
    every node/.style={font=\small}, align=center
]
    \node[block] (dev) {Developer};
    \node[block, below=of dev] (currentPrompt) {Current Prompt Template $\mathcal{P}_{i + 1}$};
    \node[block, right=of currentPrompt] (histPrompt) {Version History $\mathcal{P}_0, \hdots, \mathcal{P}_i$};
    \node[block, below=of currentPrompt] (llm) {LLM};
    \node[block, below=of llm] (response) {Output};
    \node[block, right=of llm] (examples) {All Example Inputs and Outputs};
    \node[process, above right=-0.1cm and 0.7cm of histPrompt] (generation) {\circled{1}\\Generate Assertion Candidates};
    \node[process, below=of generation] (filtering) {\circled{2}\\Filter Bad Assertions};
    \node[block, below=of filtering] (assertions) {Assertions to Deploy};

    \draw[line] (dev) -- (currentPrompt) node[midway, right] {edits};
    \draw[line] (currentPrompt) -- (llm);
     \draw[line] (currentPrompt) -- (histPrompt) node[midway, below] {saves to};
    \draw[line] (dev.west) -- (-2, 0)  |- (response.west) node[pos=1, above left] {labels};
    \draw[line] (examples) -- (llm);
    \draw[line] (llm) -- (response);
    \draw[line] (response) -| (examples) node[midway, above left] {saves to};
    \draw[line] (histPrompt) -- (generation);
    \draw[line] (examples) -- (filtering);
    \draw[line] (generation) -- (filtering);
    \draw[line] (filtering) -- (assertions);

    \begin{scope}[on background layer]
    \node[fit=(generation) (filtering) (assertions), fill=lightblue, inner sep=0.5cm, label={[anchor=north, inner sep=4pt]above:{\LARGE\(\spadesuit\) \bf \spade}}] (spade) {};
    \end{scope}

\end{tikzpicture}
\caption{Before a developer deploys a prompt template to production, \spade analyzes the deltas (i.e., diffs) between consecutive prompt templates to generate assertions. Then, \spade uses labeled pipeline inputs and outputs to filter out redundant and inaccurate assertions, while maintaining coverage of bad outputs.}
\label{fig:spade-flow}
\end{figure}

There is a lot of excitement around 
the use of large language models (LLMs)
for processing, understanding, and generating data~\cite{fernandezhowllms2023}. 
Without needing large labeled datasets,
one can easily create an {\em LLM pipeline}
for any task on a collection of data items---simply by 
crafting a natural language {\em prompt} that
instructs the LLM on what to do with
each item. 
This could span traditional data processing tasks, 
such as summarizing each document in a corpus,
extracting entities from each news article in a collection of articles,
or even imputing missing data for each tuple in a relation~\cite{narayan2022can}. 
LLMs additionally enable new and more complex data processing 
tasks that involve generating data, e.g., 
an LLM could write an explanation for why a product was recommended
to a user, author emails to potential sales leads,
or draft blog posts for marketing and outreach.
In all of these data processing tasks, deploying these LLM 
pipelines at scale---either offline, on each batch of data items, or online, as and when new items arrive---presents significant challenges, due to data quality errors made by LLMs 
seemingly at random~\cite{kalai2023calibrated}---with LLMs often
disregarding instructions, making mistakes with 
the output format, 
or hallucinating facts~\cite{zamfirescu2023johnny, shankar2022operationalizing}.

One approach to catch  
errors in deployed LLM pipelines 
is via {\em data quality assertions}.
Indeed, multiple recent papers~\cite{rebedea2023nemo, singhvi2023dspy, kim2023evallm}
and LLM pipeline authoring systems~\cite{guardrails,langchain,llama-index}
provide mechanisms to embed manually-provided or selected
assertions as part of LLM pipelines
to catch errors during deployment.  
However, determining {\em which} assertions
to add remains an open problem---and is
a big customer painpoint based on our experience
at LangChain---a company that helps people build LLM pipelines. 
Developers 
often find it difficult to determine
the right set of assertions for their custom tasks~\cite{parnin2023building}.
Challenges include predicting 
all possible LLM failure modes, 
the time-consuming nature of 
writing assertions with various 
specification methods (like Python functions or LLM calls), 
the necessity for precision in assertions 
(especially those involving LLM calls), 
and the fact that many LLM pipeline developers 
lack software engineering expertise or 
coding experience~\cite{kim2023evallm, zamfirescu2023johnny}. 
Moreover, if there are non-informative assertions
or too many of them, 
developers can get overwhelmed 
monitoring the results of these assertions.
While there is some work on 
automatically detecting
errors in traditional ML pipelines~\cite{schelterdataval, shankar2023automatic, polyzotis2018data, breck2019data},
this line of work operates
on many structured records
at a time, 
and doesn't apply to an unstructured setting.
So, we target the following question:
{\em can we identify data quality 
assertions for LLM pipelines with as little effort from developers
as possible?}

\topic{Example LLM Pipeline} 
Consider an LLM pipeline for a movie streaming 
platform, where the task is to generate a paragraph of text 
explaining why a specific movie was recommended to a specific user. 
A developer might write a prompt template 
like: {\em ``Given the following information 
about the user, \{personal\_info\}, and information about a movie, \{movie\_info\}: write a personalized note 
for why the user should watch this movie''} 
to be executed for many user-movie pairs.
In theory, this prompt seems adequate, 
but the developer might observe some data 
quality issues while testing it across different inputs: 
the LLM output might reference a movie 
the user never watched, 
cite a sensitive attribute 
(e.g., race or ethnicity), or even 
exhibit a basic issue by being too short. 
To fix these problems, developers
typically add instructions
to the prompt to catch these data quality issues,
such as ``Don't reference race in your answer''.
However, the LLM may still violate these instructions
in an unpredictable manner for some data items, requiring assertions
applied to LLM outputs post-hoc, during deployment.

\topic{Analyzing Prompt Version Histories}
To automatically synthesize assertions for developers, 
our first insight is that we can mine
prompt version histories to identify 
assertion criteria for LLM pipelines,
since developers implicitly embed
data quality requirements through changes
to the prompt, or {\em prompt deltas}, over time.
In our example above, instructions such as 
``Make sure your response is at least 3 sentences'' or ``Don't reference race in your answer''
could each correspond to a candidate assertion. 
To understand
the types of prompt deltas in LLM pipelines
and verify their usefulness for data quality assertions,
we present an 
analysis of prompt version histories of 19 custom pipelines 
from LangChain users. 
Using this analysis, we construct a taxonomy of prompt deltas (Figure~\ref{fig:taxonomy}),
which may be of independent interest for researchers and practitioners
studying how to best build LLM pipelines.

\topic{Redundancy in Data Quality Assertions}
Our second insight from analyzing these pipelines
is that if we were to create candidate
assertions from prompt deltas, 
say, using an LLM,
there may be far too many assertions---often 
exceeding 50 for just a few prompt deltas.
Many of these assertions (or equivalently, prompt deltas) are redundant, while some 
are too imprecise or ambiguous to be useful (e.g., ``return a concise response'').
Reducing this redundancy is nontrivial, 
even for engineers: assertions themselves can 
involve LLM calls with varying accuracies, 
motivating an automated component that can filter out bad candidates.
One approach is to use a small handful of 
developer-labeled LLM outputs 
to estimate each data quality assertion's 
false positive and false negative rate---but collectively
determining the right {\em set} of assertions that
can catch most errors, while incorrectly flagging 
as few outputs as possible, is not straightforward.

\topic{\spade: Our Data Quality Assertion Generation Framework} 
In this paper, 
we leverage the aforementioned
insights to develop \spade ({\bf S}ystem for {\bf P}rompt {\bf A}nalysis and {\bf D}elta-Based {\bf E}valuation, \Cref{fig:spade-flow}).
\spade's goal is to select
a set of boolean assertions 
with minimal overlap,
while maximizing coverage of bad outputs
and minimizing false failures (correct outputs
that are incorrectly flagged) for the conjunction
of selected assertions. 
We decompose \spade 
into two components---candidate assertion generation and filtering.

\subtopic{Component 1: Prompt Deltas for Candidate Assertion Generation} 
For generating candidate assertions, 
instead of directly querying 
an LLM to ``write assertions for {\em x} prompt,'' which causes
the LLM to miss certain portions of the prompt,
we generate candidates from each prompt delta, which typically 
indicate specific failure modes of LLMs. 
\spade leverages the aforementioned
taxonomy of prompt deltas we constructed
by first automatically categorizing
deltas using the taxonomy, 
then synthesizing 
Python functions (that may include LLM calls) 
as candidate assertions. 
{\bf \em At LangChain, we publicly release this 
component of \spade---which has been subsequently used 
for over 2000 pipelines across 
more than 10 sectors like 
finance, medicine, and IT~\cite{spadeblog}.}
We present an analysis of this usage
in \Cref{sec:deployment}.

\subtopic{Component 2: Filtering Candidate Assertions with Limited Data} 
To filter out incorrect and redundant candidate assertions,
instead of requiring cumbersome manual selection or even fine-tuning of separate models~\cite{saad2023ares, wang2023pandalm}, we propose an automated
approach that only requires a small handful of labeled examples, 
which are usually already present
in most target applications. 
Using these examples, we could estimate each assertion's 
false failure rate (FFR), i.e., how often an assertion incorrectly flags 
failures,
and eliminate individual assertions that exceed a given threshold. 
However, given we are selecting a set of assertions, 
the set may still exceed the FFR threshold
and flag too many failures incorrectly,  
and redundancies may persist. 
We show that selecting a small subset of 
assertions to meet failure coverage 
and FFR criteria is NP-hard. 
That said, we may express the problem 
as an integer linear program (ILP) and 
use an ILP solver to identify a solution
in a reasonable time given the size of our problem 
(hundreds to thousands of variables). 

In some cases, when there 
are limited developer-provided examples, we find that 
labeled LLM outputs may not cover 
all failure modes, leading to omission of valuable 
data quality assertions. 
For instance, in our movie recommendation scenario, 
an assertion that correctly verifies 
if the output is under 200 words 
will get discarded if all outputs in our developer-labeled sample respect 
this limit. 
To expand coverage, active learning and weak supervision 
approaches can be used to sample and label 
new LLM input-output pairs for each 
candidate assertion~\cite{castro2008active, ratner2020snorkel}, 
but this may be expensive or inaccessible for non-programmers. 
We introduce assertion {\em subsumption} as a 
way of ensuring comprehensive coverage: if one assertion 
doesn't encompass the failure modes of another, both may be selected. 
As such, \spade selects a minimal set of assertions 
that respects failure coverage, accuracy, {\em and} subsumption constraints.  

Overall, we make the following contributions:
\begin{itemize}
    \item We identify prompt version history 
    as a rich source for LLM output correctness, 
    creating a taxonomy of assertion criteria from 19 diverse LLM pipelines (\Cref{sec:candidateassertions}),
    \item We introduce \spade, our system that 
    automatically generates data quality assertions for LLM pipelines. In a public release of \spade's candidate assertion generation component\footnote{\url{https://spade-beta.streamlit.app}}, we observe and analyze 2000+ deployments (\Cref{sec:deployment}),
    \item We present a method to select a minimal set of assertions while meeting coverage and accuracy requirements, used by \spade to reduce the number of assertions. 
    For low-data settings, we introduce assertion subsumption 
    as a novel proxy for coverage (\Cref{sec:methods}), and
    \item We demonstrate \spade's effectiveness on 
    nine real-world LLM pipelines (eight of which we open-source). 
    In our low-data setting (approximately 75 inputs and outputs per pipeline), our subsumption-based solution outperforms simpler baselines that do not consider interactions between assertions by reducing the number of assertions by 14\% and lowering the false failure rate by 21\% (\Cref{sec:experiments}).
\end{itemize}

%% file: sections/candidateassertions.tex

\label{sec:candidateassertions}

\input{figures/taxonomy}

\input{figures/exampleversions}

Our first goal is to generate a set of candidate assertions. We describe how {\em prompt deltas} can inform candidate assertions and explain how to derive candidate assertions from them.

\subsection{Prompt Deltas}

A single-step {\em LLM pipeline} 
consists of a prompt template $\mathcal{P}$, 
which is formatted with a serialized 
input tuple $t$ to derive a prompt
that is fed to an LLM,  
returning a response. 
There can be many versions of $\mathcal{P}$, 
depending on how a developer 
iterates on their prompt template. 
Let $\mathcal{P}_0$ be the empty string, the 0th version, and let $\mathcal{P}_i$ be the ith version of a template. 
In the movie recommendation example from the introduction, 
suppose there are 7 versions, 
where $\mathcal{P}_7$ is the following: 
{\em ``Given the following information 
about the user, \{personal\_info\}, and information about a movie, \{movie\_info\}: write a personalized note 
for why the user should watch this movie. Ensure the recommendation note is concise, not exceeding 100 words. Mention the movie's genre and any shared cast members between the \{movie\_name\} and other movies the user has watched. Mention any awards or critical acclaim received by \{movie\_name\}. Do not mention anything related to the user's race, ethnicity, or any other sensitive attributes.''}

We define a prompt delta $\Delta \mathcal{P}_{i+1}$ 
to be the {\em diff} (or difference) 
between $\mathcal{P}_i$ and $\mathcal{P}_{i+1}$. 
Concretely, a prompt delta $\Delta \mathcal{P}$ is a set of sentences, 
where each sentence is tagged as an addition (i.e., ``+'') or deletion (i.e., ``-'').  \Cref{tab:prompt-versions} 
shows the $\Delta\mathcal{P}$s for a number of versions for our example. 
Each sentence in $\Delta\mathcal{P}_i$ is composed of additions 
(i.e., new sentences in $\mathcal{P}_i$ that didn't exist in $\mathcal{P}_{i-1}$) 
and deletions (i.e., sentences in $\mathcal{P}_{i-1}$ that don't exist in $\mathcal{P}_{i}$). 
A modification to a sentence is represented by a deletion and addition\techreport{---for example, $\Delta\mathcal{P}_{6}$ in \Cref{tab:prompt-versions} contains some new instructions added to a sentence from $\mathcal{P}_{5}$}. 
Each addition in $\Delta\mathcal{P}_{i}$ indicates possible assertion criteria, as shown in the right-most column of \Cref{tab:prompt-versions}.

\subsection{Prompt Delta Analysis}
\label{sec:taxonomy}

To understand what assertions developers may care about, we turn to real-world LLM pipelines. We analyzed 19 LLM pipelines collected 
from LangChain users, 
each of which consists of between three and 11 historical 
prompt template versions. 
These pipelines span various tasks 
across more than five domains (e.g., finance, marketing, coding, education, health), from generating workout summaries to a chatbot acting as a statistics tutor. 
\papertext{Details of the pipelines can be found in our tech report~\cite{shankar2024spade}.}
\techreport{\Cref{tab:chains} in \Cref{app:chains} shows a summary of the pipelines, including a description of each pipeline and the number of prompt versions.} 
For each pipeline, we categorized prompt deltas, i.e., $\Delta \mathcal{P}_i$, into different types---for example, instructing the LLM to include a new phrase in each response (i.e., inclusion), or instructing the LLM to respond with a certain tone (i.e., qualitative criteria). 
Two authors iterated on the categories 
4 times through a process of open and axial coding, 
ultimately producing the taxonomy in~\Cref{fig:taxonomy}. 
The taxonomy-annotated dataset of prompt versions 
can be found online\footnote{\url{https://github.com/shreyashankar/spade-experiments/blob/main/taxonomy_labels.csv}}. 

We divide deltas into two main high-level categories: 
Structural and Content-Based. Around 35\% categories identified across all deltas in our dataset were Structural, and 65\% were Content-Based.
Structural deltas indicate a minor restructuring of the prompt, without changing any criteria of a good response (e.g., adding a newline for readability),
or specification of the intended output (e.g., JSON or Markdown). 
Plausible assertion criteria based on structural deltas 
would check if the LLM output adheres to the user-specified structure. 
On the other hand, content-based deltas 
indicate a change in the {\em meaning} or definition of the task. 
Content-based deltas include descriptions of the workflow 
steps that the LLM should perform (e.g., ``first, do X, then, come up with Y''), 
instructions of specific phrases to include or exclude in responses, or
qualitative indicators of good responses (e.g., ``maintain a professional tone''). 
The Data Integration subcategory (under Content-Based deltas) 
concerns adding new sources of context to the prompt---for example, 
adding a new variable like ``\{movie\_info\}'' 
to the prompt, indicating a new type of information 
to be analyzed along with other content in the prompt. 
For some illustrative examples of prompt deltas for each category, 
we categorize the prompt deltas in \Cref{tab:prompt-versions}, 
and in \Cref{tab:movie-task}, we show sample prompt deltas 
for each category in our taxonomy. 
This taxonomy may be of independent
interest to researchers studying the process
of prompt engineering, as well as practitioners
seeking to identify ways to improve their prompts
for production LLM pipelines.


Overall, our exercise in building
this taxonomy reveals two key findings.
First, developers across diverse LLM pipelines
iterate on prompts in similar ways as encoded
as nodes in the taxonomy,
many of which correspond to 
aspects that can be explicitly checked via assertions.
Second, we find that an automated approach to
synthesize data quality assertions may be promising,
since our taxonomy reveals several such instances
where the prompt delta could directly correspond to
an assertion that captures the same requirement.
For example, many deltas correspond to instructions to 
include or exclude specific phrases, 
indicating that developers 
may benefit from assertions that explicitly verify the 
inclusion or exclusion of such phrases in LLM outputs. 

However, if we are 
to use an LLM to automatically generate assertion criteria 
based on our taxonomy of deltas, 
we also need to test whether LLMs can accurately 
identify the categories corresponding to a delta. 
Therefore, we confirmed GPT-4's correct 
categorization of prompt deltas (as of October 2023): 
we assigned ground truth categories to all 
prompt deltas from the 19 pipelines, and GPT-4 achieved an 
F1 score of 0.8135. 
\papertext{The prompt used for category extraction from prompt deltas is detailed in our tech report~\cite{shankar2024spade}.} \techreport{The prompt used for category extraction from prompt deltas is detailed in \Cref{app:prompts}.}

\begin{table}[h]
\centering
\footnotesize
\begin{tabular}{p{2.2cm}|p{2.2cm}|p{3cm}}
\toprule
\textbf{Category} & \textbf{Explanation} & \textbf{Example Prompt Delta} \\
\midrule
Response Format Instruction & Structure guidelines. & ``+ Start response with `You might like...’.'' \\
Example Demonstration & Illustrative example. & ``+ For example, here is a response for sci-fi fans...'' \\
Prompt Clarification & Refines prompt/removes ambiguity. & ``- Discuss/+ Explain movie fit...'' \\
Workflow Description & Describe ``thinking'' process. & ``+ First, analyze viewing history...'' \\
Data Integration & Adds placeholders. & ``+ Include user's \{genre\} reviews.'' \\
Quantity Instruction & Adds numerical content. & ``+ Keep note under 100 words.'' \\
Inclusion Instruction & Directs specific content. & ``+ Mention movie awards/acclaim.'' \\
Exclusion Instruction & Advises on omissions. & ``+ Avoid movie plot spoilers.'' \\
Qualitative Criteria & Sets stylistic attributes. & ``+ Maintain friendly, positive tone.'' \\
\bottomrule
\end{tabular}
\caption{Categories of prompt deltas.}
\label{tab:movie-task}
\end{table}


\subsection{From Taxonomy to Assertions}

\begin{figure}
    \centering
     \small
    \begin{tikzpicture}[auto]
    \tikzstyle{process} = [rounded corners, minimum width=7cm, text centered, text width=\linewidth]
    \tikzstyle{arrow} = [{-Latex}]

    \node (constructDelta) [process] {{\bf Construct $\Delta \mathcal{P}_5$} \\ ``- Include elements from the movie's genre, cast, and themes that align with the user's interests. + Mention the movie's genre and any shared cast members...''};
    \node (listCriteria) [process, below=0.5cm of constructDelta] {{\bf Prompt LLM with $\Delta \mathcal{P}_5$ to identify the delta type(s) and assertion criteria based on our taxonomy} \\ 
\begin{mypython}
{"criterion": "The response should mention the movie's genre.", "category": "Inclusion", "source": "Mention the movie's genre"},
{"concept": "The response should include shared cast members between the specified movie and other movies the user has watched.", "category": "Inclusion", "source": "any shared cast members between the {movie_name} and other movies the user has watched"}
\end{mypython}
    };
    \node (generateFunctions) [process, below=0.5cm of listCriteria] {{\bf Synthesize Function(s)} \\ 
\begin{mypython}
def assert_mention_of_movie_genre(ex: dict, prompt: str, response: str):
    expected_genre = ex.get('movie_genre', '').lower()
    return expected_genre in response.lower()

def assert_accurate_inclusion_of_shared_cast_members(prompt: str, response: str):
    # Formulate questions for the 'ask_llm' function to check for presence and accuracy
    presence_question = "Does the response include shared cast members between the specified movie and other movies?"
    accuracy_question = "Are the shared cast members mentioned in the response accurate and correctly representing those shared between the specified movie and other movies?"

    return ask_llm(prompt, response, presence_question) and ask_llm(prompt, response, accuracy_question)
\end{mypython}
};

    \draw [arrow] (constructDelta) -- (listCriteria);
    \draw [arrow] (listCriteria) -- (generateFunctions);

\end{tikzpicture}
    \caption{Generating candidate assertions from a $\Delta\mathcal{P}$.}
    \label{fig:stage-one}
\end{figure}

As we saw previously, prompt delta
as categorized into nodes in our taxonomy
often correspond to meaningful assertion criteria.
Next, we need
a method to automatically synthesize the data quality assertions
from the prompt deltas.
A natural idea is to prompt an LLM to generate assertion 
functions corresponding to relevant categories in our taxonomy 
given the prompt deltas. We tried this approach with GPT-4 in January 2024 
and observed several omitted assertions that clearly corresponded 
to categories in our taxonomy. Therefore, we adopt a two-step prompting process in our approach, as it has been demonstrated that breaking tasks 
into steps can enhance LLM accuracy~\cite{wei2022chain, wu2022ai, grunde2023designing}. 
In the first step, we prompt GPT-4 for natural language descriptions 
of assertion criteria, and in the second step, 
we prompt GPT-4 to generate Python functions that implement such criteria. 
\techreport{However, it's worth noting that with daily advancements in LLMs, a one-step process might already be feasible. Additionally, while our taxonomy guides LLM-generated assertions now, future LLMs may implicitly learn these categories through reinforcement learning from human feedback~\cite{christiano2017deep}. Nevertheless, knowing the taxonomy-based categories associated with candidate assertions may help in filtering them.}

More specifically, 
for each prompt delta $\Delta \mathcal{P}_i$, we first prompt 
an LLM 
to suggest as many criteria as possible for assertions---each 
aligning with a taxonomy category from \Cref{fig:stage-one}. 
A criterion is loosely defined as some natural 
language expression that operates on a given output or example
and evaluates to True or False (e.g., ``check for conciseness''). 
Our method analyzes every $\Delta \mathcal{P}_i$ instead of just the 
last prompt version for several reasons: 
developers often remove instructions from prompts 
to reduce costs while expecting the same behavior~\cite{parnin2023building}, 
prompts contain inherent ambiguities and imply multiple ways 
of evaluating some criteria, 
and complex prompts may lead to missed assertions 
if only one version is analyzed. 
Consequently, analyzing each $\Delta \mathcal{P}_i$ increases the 
likelihood of generating relevant assertions.

For each delta, our method collects the criteria identified 
and prompts an LLM again to create Python assertion functions. 
The synthesized functions can use external Python libraries 
or pose binary queries to an LLM for complex criteria. 
For function synthesis, the LLM is instructed that 
if the criterion is vaguely specified or open to interpretation, 
such as ``check for conciseness,'' it can generate multiple 
functions that each evaluate the criterion. 
In this conciseness example, the LLM could return 
multiple functions---a function that splits the response into sentences 
and ensures that there are no more than, say, 3 sentences, 
a function that splits the response into words and 
ensures that there are no more than, say, 25 words, 
or a function that sends the response to an LLM and asks whether the response is concise. 
The overall outcome of this is a multiset of candidate functions $F = \{f_1, \dots, f_m \}$. \papertext{The two prompts for generating assertion criteria and functions can be found in our tech report~\cite{shankar2024spade}.}\techreport{The two prompts for generating assertion criteria and functions can be found in \Cref{app:prompts}.}

\subsection{Initial Deployment}
\label{sec:deployment}

\begin{figure}
    \centering
    \includegraphics[width=\linewidth]{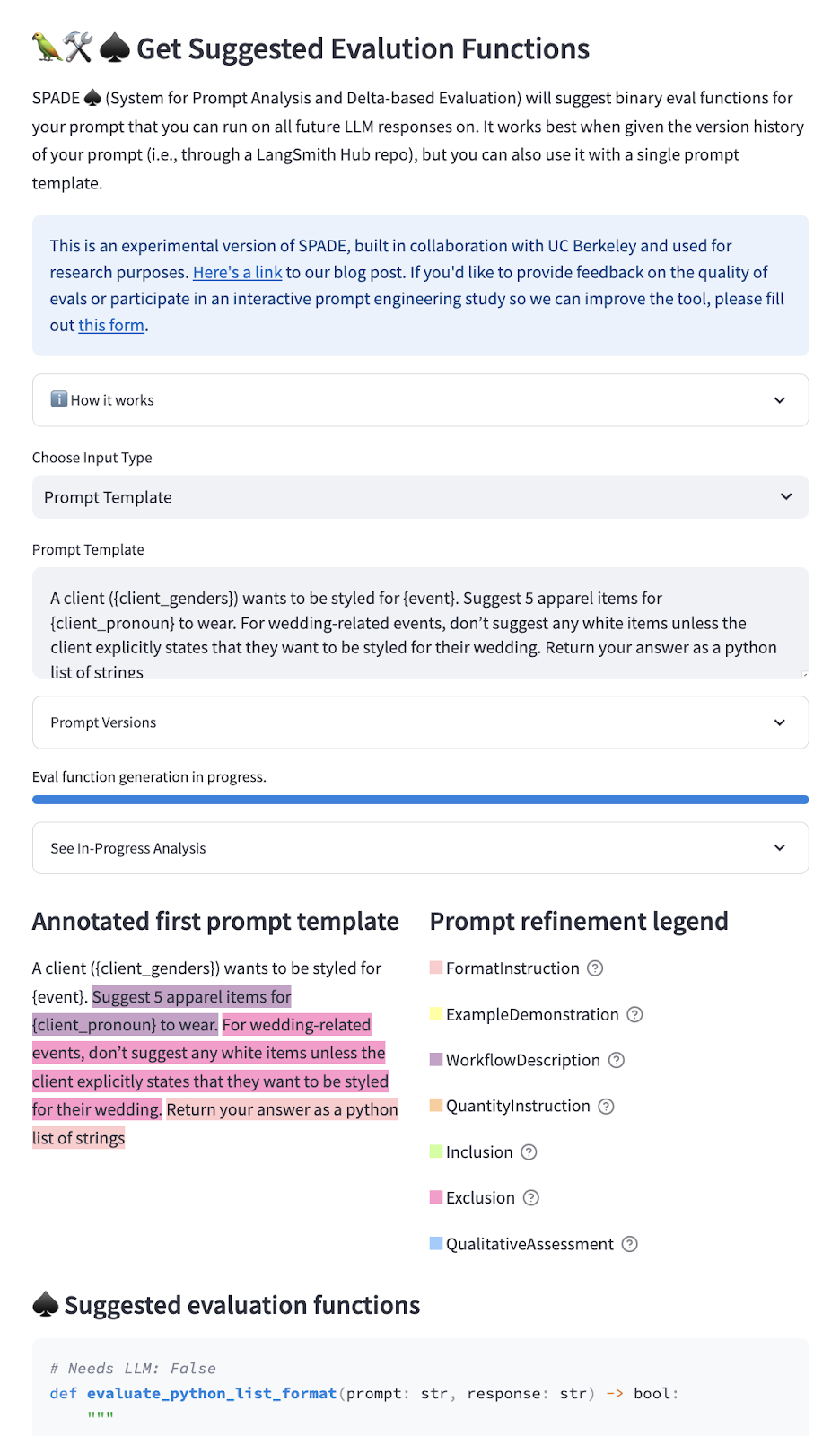}
    \caption{Screenshot of an early version of the \spade Streamlit application.}
    \label{fig:spadescreenshots}
\end{figure}

To assess the potential of our auto-generated assertions, 
in November 2023, we released an early prototype of \spade's 
candidate assertion generation framework 
via a Streamlit application\footnote{\url{https://spade-beta.streamlit.app/}}. 
\techreport{At the time of public release, the Streamlit application had a different set of prompts to generate assertions; these prompts generated relatively few assertions compared to the number of assertions generated in this paper's version of \spade. However, the taxonomy has remained the same.} 
In the Streamlit app, a developer can either paste their prompt template 
that they want to generate assertions for, 
or they can point to their LangChain Hub prompt template 
(which contains prompt version histories via commits). 
The app then visualizes the identified taxonomy categories 
in the user's prompt and displays the candidate assertions, 
as shown in the screenshot in \Cref{fig:spadescreenshots}. 

\topic{Tool Usage Insights and Feedback} 
From the reception to our Streamlit application, 
we found significant interest in auto-generated data quality assertions
for LLM pipelines: 
there have been {\em over 2000 runs} 
of the app for custom prompt templates (i.e., not the sample default prompt template in app). 
These runs span many fields, including medicine, 
education, cooking, and finance---providing insights 
from a diversity of use cases for LLM pipelines. 
\Cref{fig:streamlitqueries} shows a rough breakdown 
of the use cases people wanted to generate assertions for; 
however, it's important to note that some runs 
may not cleanly fit into a single category 
(e.g., a chatbot for telehealth-related questions 
for a medical provider could be in ``customer support'' and ``health''). 
Interestingly, 8\% of tasks related to conversational assistants, 
and we observed instances of at least 
four different companies generating assertions for their chatbots, 
given that the company name was in the prompt. 
Users for 48 of these runs clicked the ``download assertions'' button, 
which downloads the candidate assertions as a Python file. 
We note that users can also directly copy the assertions code displayed, 
instead of downloading the assertions as a Python file, 
and we did not measure the copy events. 
No users clicked the ``thumbs down'' button---which is not to say 
that the generated assertions were perfect. \techreport{In fact, anecdotally, we have seen the candidate assertion quality improve over the last several months as GPT-4 continually improves. On average, 3.3 assertions were generated per run, with minimum 1 and maximum 10 assertions generated for a run. Most of the runs corresponded to a single prompt version.}
After our Streamlit app was released, 
we found an unprompted review of 
\spade-generated assertions from a LangChain user who built a 
``chat-with-your-pdf'' tool\footnote{\url{https://twitter.com/th_calafatidis/status/1728144652119769394}}: in their words, 
{\em ``When I saw it I didnt beleive it could work that well, but it really did and made the evaluation process fun and ez.''} 
\papertext{We also found an independently-written 
open-source re-implementation of \spade\footnote{\url{https://github.com/uptrain-ai/uptrain/blob/main/examples/integrations/spade/evaluating_guidelines_generated_by_spade.ipynb}}.}

\topic{Observations about Assertion Criteria} 
Across the 2000+ runs, 
with regards to our taxonomy, assertion criteria 
were most commonly derived from inclusion and exclusion instructions. 
For example, for a shopping assistant, where the objective 
was to find the most relevant product related to a customer's query, 
responses were required to include ``all the features [the customer] asked for.'' 
In another customer support agent example, responses were required to 
include ``exact quotes in the [context] relevant to the [customer's] question...word for word.'' 
One common exclusion instruction across chat-related tasks 
was to avoid any discussion unrelated to the end-user's question; 
however, we noticed that GPT-4 struggled to generate assertion criteria 
around such a generic exclusion instruction. 
We found Python functions generated to 
implement criteria such as ``avoid unrelated ideas'',  
did not use an LLM to check the criteria. 
Instead, they checked for the presence of specific phrases like 
``unrelated'' and ``I'm sorry, I did not find anything related'' 
in the response. 
Such errors indicate the limitations of current 
state-of-the-art LLMs and may suggest the need for 
specialized, fine-tuned LLMs for generating assertions in future work;
here, we work with the limitations of present-day state-of-the-art LLMs. 
\papertext{We discuss this further in our tech report~\cite{shankar2024spade}.} \techreport{We discuss this further in \Cref{sec:experiments-limitations-future-work}.} 

A new pattern we noticed after deployment, 
which wasn't seen in our initial analysis, 
is that developers often wish to hide 
certain parts of the LLM workflow in the outputs. 
This could be viewed as a special instance of the 
exclusion category in our taxonomy. 
For instance, in a prompt related to enterprise automation, 
the first instruction was to write a query targeting 
a specific database (e.g., ``Write a SQL query to 
fetch the most relevant table from the MySQL database''). 
However, another instruction in the prompt specified 
that the name of the database should not appear in the final summary 
returned to the end-user (e.g., ``Do not mention that you queried the MySQL table X''). 
We were pleasantly surprised to find that our process for generating criteria 
successfully identified and implemented such exclusion instructions accurately.

\begin{figure}
    \centering

\resizebox{\linewidth}{!}{\begin{tikzpicture}
\begin{axis}[
    xbar,
    xlabel={Percentage},
    symbolic y coords={
        Health,
        IT and Programming,
        Project Management,
        Customer Support,
        Education,
        Sports and Entertainment,
        Cooking,
        Finance,
        Marketing,
        Others},
    ytick=data,
    y tick label style={align=right, text width=3cm}, 
    nodes near coords,
    nodes near coords align={horizontal},
    xmin=0, xmax=24,
    ]

\addplot coordinates {
    (20.8,Health) 
    (16.7,IT and Programming) 
    (12.5,Project Management) 
    (12.5,Customer Support) 
    (8.3,Education) 
    (8.3,Sports and Entertainment) 
    (8.3,Cooking) 
    (4.2,Finance) 
    (4.2,Marketing) 
    (4.2,Others)};
\end{axis}
\end{tikzpicture}}

    \caption{Prompts submitted to the \spade Streamlit application span a variety of fields.}
    \label{fig:streamlitqueries}
\end{figure}

\topic{Observations about Assertions} 
In LLM pipelines with numerous prompt versions, 
we observed two main patterns. 
First, prompt engineering often leads to many similar assertions, 
and redundancy can be a headache at deployment 
when a developer has to keep track of so many assertions. 
For instance, a pipeline to summarize 
lecture transcripts\footnote{\url{https://smith.langchain.com/hub/kirby/simple-lecture-summary}} 
had 14 prompt versions, with many edits 
localized to the paragraph providing instructions on titles, 
speakers, and dates, creating multiple overlapping assertions. 
10 assertions assessed the lecture's 
central thesis; two are as follows: 

\begin{mypython}
async def assert_response_articulates_central_thesis(
    example: dict, prompt: str, response: str
):
    return "Central Thesis:" in response

async def assert_response_completeness(example: dict, prompt: str, response: str):
    required_elements = [
        "Context:",
        "Central Thesis:",
        "Key Points:",
        "Conclusions and Takeaways:",
        "Glossary of Important Terms:",
    ]
    return all(element in response for element in required_elements)
\end{mypython}

Second, many assertions may be incorrect, 
causing runtime errors or false failures. 
Assessing the accuracy of these assertions is challenging, 
particularly for those that are complex or invoke LLMs themselves, 
where even experienced developers might not be able to 
gauge their effectiveness without viewing 
the results of the functions on many examples. 
Even if the assertions are not incorrect, they still may
have undesirable failure rates or coverage,
which is often hard for a developer to reason about in
conjunction with other assertions. 
Since there can be 50+ assertions generated 
for just a handful of prompt versions 
(as demonstrated in \Cref{sec:experiments}), 
manually filtering them by eyeballing failure rates for each subset of assertions is impractical. 
Therefore, we adopt an automated approach for filtering,
as discussed in the following section.

%% file: figures/taxonomy.tex
\begin{figure}
    \centering
\includegraphics[width=\linewidth]{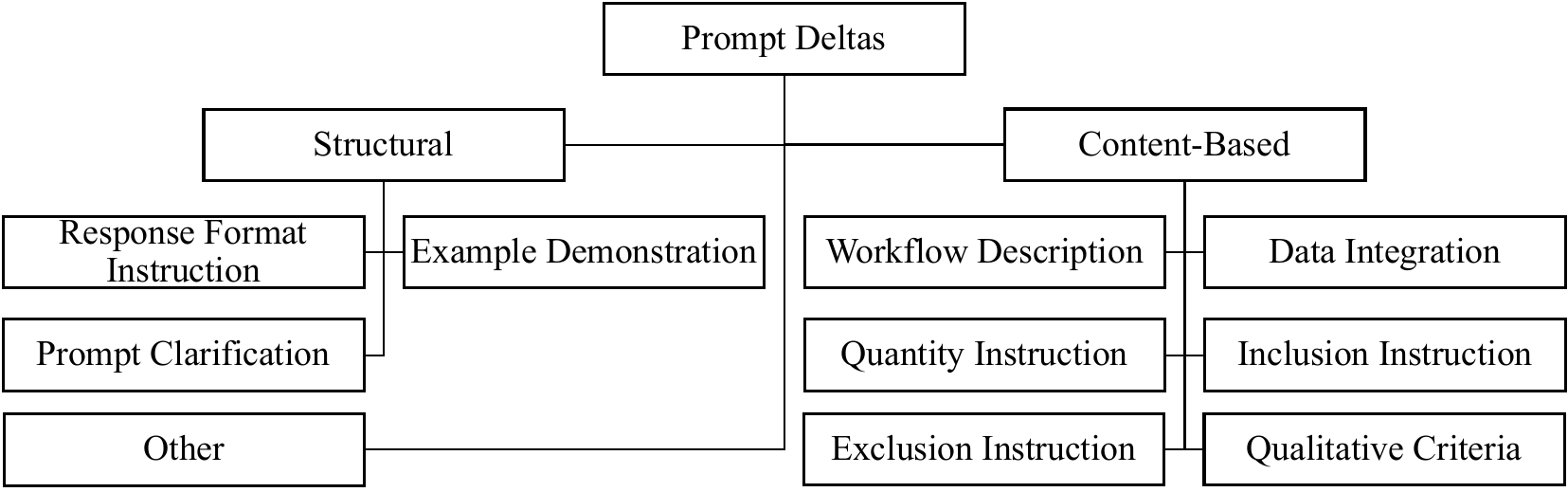}
    \caption{Taxonomy of prompt deltas from 19 LLM pipelines.}
    \label{fig:taxonomy}
\end{figure}

%% file: figures/exampleversions.tex
\begin{table*}[h]
\centering
\small
\begin{tabularx}{\textwidth}{c|X|l|X}
\toprule
\textbf{Version $i$} & \textbf{$\Delta\mathcal{P}_i$} & \textbf{$\Delta\mathcal{P}_i$ Category} & \textbf{Possible New Assertion Criteria} \\
\midrule
1 & + Given the following information about the user, \{personal\_info\}, and information about a movie, \{movie\_info\}: write a personalized note for why the user should watch this movie. & Inclusion & Response should be personalized and relevant to the given user information \\

2 & + Include elements from the movie's genre, cast, and themes that align with the user's interests. & Inclusion & Response includes specific references to the user's interests related to the movie's genre, cast, and themes \\

3 & + Ensure the recommendation note is concise. & Qualitative Assessment & Response should be concise \\

4 & - Ensure the recommendation note is concise. + Ensure the recommendation note is concise, not exceeding 100 words. & Count & Response should be within the 100 word limit \\

5 & - Include elements from the movie's genre, cast, and themes that align with the user's interests. + Mention the movie's genre and any shared cast members between the \{movie\_name\} and other movies the user has watched. & Inclusion & Response should mention genre and verify cast members are accurate \\

6 & + Mention any awards or critical acclaim received by {movie\_name}. & Inclusion & Response should include references to awards or critical acclaim of the movie \\

7 & + Do not mention anything related to the user's race, ethnicity, or any other sensitive attributes. & Exclusion & Response should not include references to sensitive personal attributes \\

\bottomrule
\end{tabularx}
\caption{Comparison of 7 prompt versions for an LLM pipeline to write personalized movie recommendations. In each $\Delta\mathcal{P}_i$, a sentence starts with ``+'' if it is a newly added sentence; a sentence starts with ``-'' if it is removed from $\mathcal{P}_{i-1}$. For each $\Delta\mathcal{P}_i$, we list its category (using the taxonomy in \Cref{fig:taxonomy}) and possible new assertion criteria.}
\label{tab:prompt-versions}
\end{table*}

%% file: sections/methods.tex

\label{sec:methods}

As we saw in the previous section, 
we often have redundant and incorrect assertions, 
particularly in pipelines with numerous prompt versions. 
Here, we focus on filtering this candidate set to a smaller number, which not only improves efficiency 
and reduces cost when deploying the assertions to run in production, but it also reduces cognitive overhead for the developer.

\subsection{Definitions}
\label{sec:methods-defs}

Consider $e_i$ as an example end-to-end execution (i.e., run) of an LLM pipeline on some input. For the purposes of this section, we assume the prompt template $\mathcal{P}$ is fixed to be the final prompt version that the developer eventually decided on. We denote $y_i \in \{0, 1\}$ to represent whether the developer considers $e_i$ to be a {\em success} (1) or {\em failure} (0). 

Let $E$ be the set of all such example runs (this set is not provided upfront, as we will deal with shortly). We define an assertion function $f: E \rightarrow \{0, 1\}$, where $1$ indicates that
the example was processed successfully by the LLM and $0$ otherwise. Let $F' = \{f_1, f_2, \ldots, f_k\}$ be a set of $k$ data quality assertions. An example $e_i$ is deemed {\em successful by $F'$} if and only if it satisfies all assertions in $F'$. Specifically:
\[\hat{y}_i = 
\begin{cases} 
1~~  \text{if } f(e_i) = 1 \quad \forall f \in F', & \text{ ~~~~ // Deemed successful by all of } F'\\
0~~  \text{otherwise.} &\text{ ~~~~ // Deemed a failure by} \geq 1 \text{ of } F'
\end{cases}\]
Given all $m$ candidate assertions $F = \{f_1, f_2, \ldots, f_m\}$, the objective is to select $F' \subseteq F$ such that $\hat{y}_i = y_i$ for most examples in $E$, with $F'$ being as small as possible. This goal involves maximizing failure coverage and minimizing the false failure rate and selected function count, expressed as follows:

\begin{definition}
\label{def:coverage}
    Coverage for a set $F'$ is the proportion of actual failures that are correctly identified by $F'$, defined as:
    \[
    \text{Coverage}\left(F'\right) = \frac{\sum_{i} \mathbb{I}\left[\hat{y}_i = 0 \land y_i = 0\right]}{\sum_{i}  \mathbb{I}\left[y_i = 0\right]} \text{ ~~~~ // Failures caught by } F'
    \]
\end{definition}

\begin{definition}
\label{def:ffr}
    False Failure Rate (FFR) for a set $F'$ is the fraction of examples that $F'$ incorrectly evaluates as failures ($\hat{y}_i = 0$) when they are actually successful ($y_i = 1$), defined as:
    \[
    \text{FFR}\left(F'\right) = \frac{\sum_{i} \mathbb{I}\left[\hat{y}_i = 0 \land y_i = 1\right]}{\sum_{i} \mathbb{I}\left[y_i = 1\right]} \text{ ~~~~ // Non-failures flagged by } F'
    \]
\end{definition}
In both definitions above, \(\hat{y}_i\) represents set $F'$'s prediction for the \(i\)-th example, \(y_i\) is the actual outcome, while  $\mathbb{I}$ is the indicator function. In practice, coverage and FFR are impossible to compute since the universe of examples $E$ is unknown. So, for now, we assume access to a subset $E' \subset E$ of labeled LLM responses, where $E'$ is a manually provided set of example runs and may not contain all the types of failures the LLM pipeline could observe---an issue we will deal with in \Cref{sec:methods-subsumptionproblem}. Thus, we replace~\Cref{def:coverage} and \Cref{def:ffr} with $\text{Coverage}_{E'}\left(F'\right)$ and $\text{FFR}_{E'}\left(F'\right)$, omitting the subscript $E'$ for brevity. \techreport{\Cref{tab:notation} in \Cref{app:notation} summarizes the notation used throughout this section.}

\subsection{Coverage Problem Formulation}
\label{sec:methods-covproblem}

Our goal is therefore to select a minimal set of assertions $F' \subseteq F$ based on a sample $E' = \{e_1, \hdots, e_n \} \subset E$. Formally:
{\begin{flalign}
& \text{minimize} \quad |F'| && \nonumber \\
& \text{subject to:} \quad \text{Coverage}(F') \geq \alpha, \quad \text{FFR}(F') \leq \tau &&\nonumber
\end{flalign}}
The above problem may be infeasible with certain values of $\alpha$, for multiple reasons. To see this, consider the case where no assertion catches a specific failure, while $\alpha$ is set to $1$. We ignore this (and similar cases) for now and defer their discussion to \Cref{sec:experiments}. To expand out the above, we introduce a matrix $M$ (size $n \times m$) to track each assertion's result on each example $e_i$, where $M_{ij} = 1$ if $f_j(e_i) = 1$ (i.e., $F_j$ deems $e_i$ a success) and $M_{ij} = 0$ otherwise. We also define binary variables $x_j$ and $w_{ij}$ to represent whether an assertion is chosen and if it marks an example as a failure: $w_{ij} = \left(1 - M_{ij}\right) \cdot x_j$, which is based on whether $f_j$ denotes $e_i$ as a failure and $f_j$ is included in $F'$. We additionally introduce the binary variable $u_i$ to represent whether a failed example is covered by any selected assertion:
{\begin{flalign*}
& u_i \leq \sum_{j=1}^{m} w_{ij}, \quad \forall i \in \left[1, n\right]: y_i = 0. \text{~~~~~~~~// Coverage of failure } e_i
\end{flalign*}}
Then, the coverage constraint can be written as:
\begin{flalign*}
& \frac{\sum_{i: y_i = 0} u_i}{\sum_i \mathbb{I} \left[y_i = 0\right]} \geq \alpha.
\end{flalign*}
\techreport{Now, we formulate the FFR constraint. Observe that FFR can be decomposed into the following:

{\setlength{\abovedisplayskip}{0pt}
\begin{flalign}
\frac{\sum_{i=1}^{n} y_i \cdot \max_j\left( w_{ij}\right)}{\sum_{i=1}^{n} \left[y_i = 1\right]} \leq \tau. \label{eq:ffrilp1}
\end{flalign}}

The product $w_{ij} = \left(1 - M_{ij}\right) \cdot x_j$ in the numerator is 1 if $f_j$ is included in $F'$ and incorrectly marks a good example as a failure. The $\max$ function ensures that as long as there is some selected function $f_j$ that marks $e_i$ as a failure, the product is 1. Multiplying $y_i$ by the $\max$ result ensures that the numerator is incremented only when $y_i = 1$, or when the example $e_i$ is actually not a failure. This numerator is thus equivalent to the numerator in \Cref{def:ffr}, i.e., $y_i = 1 \land \hat{y}_i = 0$. The entire fraction is less than the threshold $\tau$, which represents the maximum allowed false failure rate across the examples. To formulate FFR according to \eqref{eq:ffrilp1},}%
\papertext{FFR is decomposed similarly:} we introduce a new binary variable $z_i$, which defines whether $F'$ denotes $e_i$ as a failure while $e_i$ is actually a successful example (i.e., false failure):
{\begin{flalign*}
& z_i \geq y_i \cdot w_{ij}, \quad \forall i \in \left[1, n\right]; \, \forall j \in \left[1, m\right]. \text{~~~~~~~~// } e_i \text{ is a false failure}
\end{flalign*}
Then, the FFR constraint is:
{\begin{flalign*}
& \frac{\sum_{i=1}^{n} z_i}{\sum_{i=1}^{n} \mathbb{I} \left[y_i = 1\right]} \leq \tau.
\end{flalign*}}

We can then state the problem of minimizing the number of assertions while meeting $E'$ coverage and FFR constraints as an Integer Linear Program (ILP):
{\begin{flalign*}
& \text{minimize} \quad \sum_{j=1}^{m} x_j && \\
& \text{subject to:} \quad w_{ij} = (1 - M_{ij}) \cdot x_j, \quad \forall i \in \left[1, n\right], \ \forall j \in \left[1, m\right]; && \\
& \quad u_i \leq \sum_{j=1}^{m} w_{ij}, \quad \forall i \in \left[1, n\right] \ \text{where} \ y_i = 0; \quad \frac{\sum_{i: y_i = 0} u_i}{\sum_i \mathbb{I} \left[y_i = 0\right]} \geq \alpha; && \\
& \quad z_i \geq y_i \cdot w_{ij}, \quad \forall i \in \left[1, n\right], \ \forall j \in \left[1, m\right]; \quad \frac{\sum_{i=1}^{n} z_i}{\sum_{i=1}^{n} \mathbb{I} \left[y_i = 1\right]} \leq \tau; && \\
& \quad x_j, u_i, z_i, w_{ij} \in \{0, 1\}, \quad \forall i \in \left[1, n\right], \ \forall j \in \left[1, m\right]. &&
\end{flalign*}}
We refer to a solution for this ILP as $\spade_{\text{cov}}$. Trivially, the problem is NP-hard for $\tau=0$ and $\alpha=1$, via a simple reduction from set cover, and is in NP, since it can be stated in ILP form. In our case, given tens of candidate assertions and fewer than 100 examples $e_i$, the ILPs tend to be of reasonable size (i.e., thousands of variables). Most ILP solvers can efficiently and quickly such programs.

\subsection{Subsumption Problem Formulation}
\label{sec:methods-subsumptionproblem}

So far, we've assumed that the developer is willing to provide a comprehensive set of labeled example runs $E'$. In settings where the developer is unwilling to do so, and where $E'$ does not include all failure types in $E$, $\spade_{\text{cov}}$ may overlook useful assertions in $F$ that only catch failures in $E \setminus E'$---as shown empirically in \Cref{sec:experiments}. We initially considered using active learning~\cite{castro2008active} to sample more LLM responses for each assertion and weak supervision to label the responses~\cite{ratner2020snorkel}. However, this approach can be costly with state-of-the-art LLMs, and it demands significant manual effort to balance failing and successful examples for each assertion, ensuring meaningful FFRs, and avoiding the exclusion of assertions due to underrepresented failure types. For this setting, we additionally introduce the notion of {\em subsumption}. Assuming that all candidate assertion functions cover as many failure modes as possible, our goal is to pick $F' \subseteq F$ such that assertions in $F \setminus F'$ are {\em subsumed} by $F'$. Formally, a set of functions $S$ subsumes some function $f$ if the conjunction of functions in $S$ logically implies the conjunction of functions in $S$ and $f$. That is,

\begin{definition}
\label{def:subsumption}
    A set of functions $S \implies f$ if and only if $\forall e \in E, \exists s \in S \text{ such that } s (e)  \implies f (e)$. In other words, if $S \implies f$, then $f$ catches no new failures that $S$ does not already catch.
\end{definition}
A simple example of subsumption is as follows: suppose our set of functions $S$ contains only one function $f$, which parses an LLM output into a JSON list to check that the list has at least 2 elements. $f$ may be generated as a result of a count-related instruction from our taxonomy of prompt deltas. Now, let $g$ be some other function that only checks if the output can be turned into a JSON list. $g$ might have been generated due to a ``Response Format Instruction''-type delta. If $g$ fails for some LLM output, $f$---and therefore $S$---must also fail that output; thus $S \implies g$.


\subsubsection{ILP with Subsumption Constraints}
\label{sec:methods-subsumptionconstraints}

We reformulate the problem with subsumption constraints. Let $G$ be the set of functions in $F \setminus F'$ not subsumed by $F'$:
{\begin{flalign}
& \text{minimize} \quad |F'| + |G| && \nonumber \\
& \text{subject to:} \quad \text{Coverage}(F') \geq \alpha, \quad \text{FFR}(F') \leq \tau \nonumber
\end{flalign}}
To represent $G$, we introduce binary variables and a matrix $K$ to denote subsumption relationships. $K_{ij} = 1$ if and only if ${f_i} \implies f_j$. We discuss how we construct $K$ in more detail in \Cref{sec:methods-evalsubsumption}. Recall that $x_j$ represents whether \( f_j \) is selected in \( F' \). For each function $f_j$, a binary variable $r_j$ indicates if it is subsumed by $F'$.
{\begin{flalign*}
& x_i \cdot K_{ij} \leq r_j \leq \sum_{\substack{i = 1 \\ i \neq j}}^{m}  \left(x_{i} \cdot K_{ij}\right), \quad \forall j \in \left[1, m\right], i \neq j. 
\end{flalign*}}

Now, $s_j$ will denote if $f_j$ is neither in $F'$ nor subsumed by $F'$. We break this into three parts. 
First,
\begin{flalign*}
& s_j \leq 1 - x_j,  \quad \forall j \in \left[1, m\right],
\end{flalign*}
indicates that $f_j \not \in F'$ if $s_j$ is allowed to be 1 (i.e., $f_j \not \in F'$ when $x_j = 0$). The second constraint,
\begin{flalign*}
& s_j \leq 1 - r_j,  \quad \forall j \in \left[1, m\right],
\end{flalign*}
captures the condition where $f_j$ is not subsumed by $F'$. Here, if $r_j = 0$, suggesting no subsumption, then $s_j$ may be 1. Lastly, to combine these conditions, we employ the constraint,
\begin{flalign*}
& s_j \geq 1 - x_j - r_j,  \quad \forall j \in \left[1, m\right],
\end{flalign*}
which ensures that $s_j$ can only be 1 if both prior conditions are satisfied: $f_j$ is neither in $F'$ nor subsumed by $F'$. 

Finally, our objective is to minimize the sum of the number of functions in $F'$ and non-subsumed functions $G$. The ILP formulation then becomes (with changes highlighted in \textcolor{darkblueilp}{blue}):
{\begin{flalign*}
& \text{minimize} \quad \sum_{j=1}^{m} x_j + \textcolor{darkblueilp}{\sum_{j=1}^{m} s_j} && \\
& \text{subject to:}  \quad w_{ij} = \left(1 - M_{ij}\right) \cdot x_j, \ \forall i \in \left[1, n\right], \ \forall j \in \left[1, m\right]; && \\
& \quad u_i \leq \sum_{j=1}^{m} w_{ij}, \quad \forall i \in \left[1, n\right] \ \text{where} \ y_i = 0; \quad \frac{\sum_{i: y_i = 0} u_i}{\sum_i \mathbb{I} \left[y_i = 0\right]} \geq \alpha; && \\
& \quad z_i \geq y_i \cdot w_{ij}, \quad \forall i \in \left[1, n\right], \forall j \in \left[1, m\right]; \quad \frac{\sum_{i=1}^{n} z_i}{\sum_{i=1}^{n} \mathbb{I} \left[y_i = 1\right]} \leq \tau; && \\
& \quad \textcolor{darkblueilp}{x_i \cdot K_{ij} \leq r_j \leq \sum_{\substack{i = 1 \\ i \neq j}}^{m} \left(x_{i} \cdot K_{ij}\right), \quad \forall j \in \left[1, m\right], i \neq j;} && \\
& \quad \textcolor{darkblueilp}{s_j \leq 1 - x_j, \quad s_j \leq 1 - r_j, \quad \forall j \in \left[1, m\right];}  && \\
& \quad \textcolor{darkblueilp}{s_j \geq 1 - x_j - r_j, \quad \forall j \in \left[1, m\right];} && \\
& \quad x_j, u_i, z_i, w_{ij}, \textcolor{darkblueilp}{r_j}, \textcolor{darkblueilp}{s_j}  \in \{0, 1\}, \quad \forall i \in \left[1, n\right], \ \forall j \in \left[1, m\right]. && 
\end{flalign*}}
We call a solution to this ILP $\spade_{\text{sub}}$. We maintain the coverage constraint because the subsumption approach alone does not inherently account for the distribution or the significance of different types of failures. For instance, if a particular type of failure makes up a critical portion of $E'$, a subsumption-based approach might overlook it. To see this in the simplest case, consider $\alpha = 1$: simply optimizing for the sum $|F'| + |G|$ does not guarantee all failures in $E'$ are covered. In practice, $\spade_{\text{sub}}$ is less sensitive to $\alpha$ than $\spade_{\text{cov}}$, as we will discuss further in \Cref{sec:experiments}.

\subsubsection{Assessing Subsumption} \label{sec:methods-evalsubsumption} 
Here, we detail how to construct $K$, our matrix representing subsumption relationships between pairs of functions $f_i, f_j$. For pure Python functions, one could use static analysis to determine subsumption. However, it becomes complex when dealing with assertions that include LLM calls or a mix of pure Python and LLM-invoking assertions. For these, \spade employs GPT-4 to identify potential subsumptions $\{a\} \implies b$ for pairs of functions $a, b$. For any pipeline, there are only two calls to the LLM to determine all subsumptions: first, all assertion functions are combined into a single prompt for GPT-4, instructing it to list as many subsumption relationships as it can identify, then prompting it again to transform its response into a parse-able list of pairs $a \implies b$. \techreport{Details of this LLM subsumption prompt are in \Cref{app:prompts}.}



To maximize precision of $\implies$ relationships identified, we employ some heuristics. First, $E'$ can filter subsumptions: for $f_i$ and $f_j$, observe that:
{\begin{align*}
    &\exists e_i \in E' : \left( f_i\left(e_i\right) = 1\right) \land \left(f_j\left(e_i\right) = 0\right) \Rightarrow \left(\left\{\hdots,f_i,\hdots\right\} \notimplies f_j\right).
\end{align*}}
In other words, any set containing $f_i$ definitely does not subsume $f_j$ if $f_j$ flags a failure that $\{f_i\}$ does not. Next, we use the FFR threshold to skip evaluating subsumption. Observe that, for any set of assertions $S$ and $f \not \in S$,
{\begin{flalign}
    \max\left(\text{FFR}\left(S\right), \text{FFR}\left(\{f\}\right)\right) &\leq \text{FFR}\left(S \cup \{f\}\right), \nonumber \\
    \text{FFR}\left(S \cup \{f\}\right) &\leq \text{FFR}\left(S\right) + \text{FFR}\left(\{f\}\right). \label{eq:ffrbound}
\end{flalign}}
As such, we need not evaluate $\left\{f_i\right\} \implies f_j$ if either $\text{FFR}\left(\{f_i\}\right) \geq \tau$ or $\text{FFR}\left(\{f_j\}\right) \geq \tau$. Lastly, we use transitivity of implication to further prune checks: if ${x} \implies y$ and ${y} \implies z$, then ${x} \implies z$ is also true.

\subsubsection{Subsumption without Examples}
For completeness, we also consider the case where we have no 
developer-provided examples $E'$.
In this case, we are only reliant on our subsumption relationships to pick our set $F'$.
Our problem can then be restated as follows:
{\begin{flalign*}
& \text{minimize} \quad \sum_{j=1}^{m} x_j + \sum_{j=1}^{m} s_j && \\
& \text{subject to:}  \quad  x_i \cdot K_{ij} \leq r_j \leq\sum_{\substack{i = 1 \\ i \neq j}}^{m} (x_{i} \cdot K_{ij}), \quad \forall j \in \left[1, m\right]; && \\
& \quad s_j \leq 1 - x_j, \quad s_j \leq 1 - r_j, \quad \forall j \in \left[1, m\right];  && \\
& \quad s_j \geq 1 - x_j - r_j, \quad \forall j \in \left[1, m\right]; && \\
& \quad x_j, r_j, s_j  \in \{0, 1\}, \quad  \forall j \in \left[1, m\right]. && 
\end{flalign*}}
However, it turns out this problem is no longer NP-Hard.
Consider the example in Figure~\ref{fig:subsumption}, where an edge indicates subsumption.
For example $a \rightarrow b$ means that $b$ catches a subset of the failures
of $a$,
and is therefore subsumed by $a$.
Here, if we were to minimize  $\Sigma_j (x_j + s_j)$,
we would simply pick $a$ and $e$ to be part of $F'$, 
since the rest of the functions
would be subsumed (therefore their $s_j = 0$---recall that
$s_j=1$ for assertions that are neither subsumed nor selected), 
and our overall metric would evaluate to $2$.
One intuitive way to view this problem 
is to start by keeping all of the nodes as unselected,
i.e., $x_j = 0, s_j = 1$, and then by adding them one at a time 
in a predefined order to $F'$,
we set $x_j = 1, s_j = 0$, and at the same time, we also
impact $s_j$ (moving them from $1$ to $0$) 
for all other functions that then become subsumed as a result.
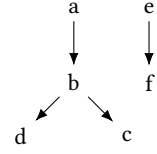
\begin{figure}
\centering
\begin{tikzpicture}[>=Latex, node distance=1cm]
    \node (a) {a};
    \node[below of=a] (b) {b};
    \node[below left of=b] (d) {d};
    \node[below right of=b] (c) {c};
    \node[right of=a] (e) {e};
    \node[below of=e] (f) {f};
    
    \draw[->] (a) -- (b);
    \draw[->] (b) -- (c);
    \draw[->] (b) -- (d);
    \draw[->] (e) -- (f);
\end{tikzpicture}
\caption{Example depicting subsumption relationships.}\label{fig:subsumption}
\end{figure}

More generally, subsumption relationships can be represented in the form of a Directed Acyclic Graph (DAG),
as in our example above.
(We omit the trivial case where two or more functions are equivalent,
that is the only case where there can be cycles;
in all other cases we have a DAG.)
Within this DAG, we simply select all nodes
that have no incoming edges and add them to $F'$.
We can see that the rest of the nodes
are subsumed, since they have at least one incoming edge. 
We could certainly exclude a node that doesn't have any incoming edge from $F'$,
but adding it to $F'$ does not worsen the objective
because there is no way that that node will be subsumed by
another.
So, overall, we need to do a topological sort of the subsumption graph
and pick the nodes ``at the top''.
Thus, the problem is in PTIME when there are no examples provided.

%% file: sections/experiments.tex

\label{sec:experiments}

\input{figures/datasetdesc}
\input{figures/spaderesults}

We first discuss the LLM pipelines and datasets (i.e., $E'$); 
then, we discuss methods and metrics and present our results. The experiment code, datasets, and LLM responses are hosted on GitHub\footnote{\url{https://github.com/shreyashankar/spade-experiments}}.

\subsection{Pipeline and Dataset Descriptions}

We evaluated \spade on nine 
LLM pipelines---eight from LangChain Hub\footnote{\url{https://smith.langchain.com/hub}}, an open-source collection of LLM pipelines, and 1 proprietary pipeline. Each pipeline consists of a prompt template for the LLM and collection of approximately 75 examples (i.e., inputs and ouptuts in $E'$) with labels for whether the outputs were good or bad. All nine pipelines come from LangChain users. Six pipelines were used in developing our prompt delta taxonomy (\Cref{sec:taxonomy}), each representing a different domain (e.g., programming, finance, marketing). Two pipelines came from \spade's Streamlit deployment (\Cref{sec:deployment}). We include one final proprietary pipeline (the {\em fashion} pipeline) in our experiments because it had the largest number of prompt template versions and \spade's assertions have already been deployed in production for tens of thousands of daily runs.

\Cref{tab:dataset-desc} provides details on each LLM pipeline and corresponding set of good and bad examples. For the {\em fashion} pipeline, good and bad examples were provided by a developer at the corresponding startup. While we used real user-provided prompt templates and histories (between 3 and 16 prompt versions) for the other 8 pipelines, we constructed and annotated our own input-output examples so we could release them publicly. For two pipelines, we sourced examples from Kaggle. For the other six pipelines, we synthetically generated the other datasets using Chat GPT Pro (based on GPT-4) and manually reviewed them. For instance, for the {\em codereviews} pipeline that uses an LLM to review pull requests, we asked Chat GPT to generate example pull requests covering a variety of programming languages, application types, and diff sizes. On average, we collected 75 examples per pipeline. We then executed the LLM pipelines on these inputs and manually labeled the responses to assess whether they met the prompt instructions.

\subsection{Method Comparison and Metrics}

As before, let $E'$ be a dataset of example prompt-response pairs, as well as the corresponding labels of whether the response was good (i.e., 1) or bad (i.e., 0). Let $\tau$ be the FFR threshold and $F$ be the set of candidate assertions produced by the first step of \spade (\Cref{sec:candidateassertions}). If a candidate function $f$ results in a runtime error for some example $e$, we denote $f\left(e\right) = 0$ (i.e., failure). All of our code was written in Python, using the PuLP Python package to find solutions for the ILPs. We used the default PuLP configuration, which uses the CBC solver~\cite{forrest2005cbc}. 

The simplest baseline involves generating candidate assertions and choosing all of them, but it proved ineffective, yielding 100\% coverage and 100\% FFR due to at least one assertion failing all tests. Failures were due to errors or overly specific conditions that, in theory, could pass some outputs (for example, requiring a precise phrase in a certain case) but, in reality, never matched any actual LLM outputs. As such, we evaluated two versions of \spade against a baseline that simply filters candidate assertions that individually exceed the FFR threshold:

\begin{itemize}
    \item $\textsc{baseline}$ selects all functions $f$ in $F$ where $\text{FFR}\left(\{f\}\right) \leq \tau$
    \item $\spade_\text{cov}$ is a solution to the ILP defined in \Cref{sec:methods-covproblem}
    \item $\spade_\text{sub}$ is a solution to the ILP defined in \Cref{sec:methods-subsumptionconstraints}
\end{itemize}

Let $F'$ represent the set of selected assertions by any version of \spade. We measure four metrics:

\begin{enumerate}
    \item Fraction of Assertions Selected (i.e., $|F'|/|F|$)
    \item Fraction of Excluded Non-Subsumed Functions (i.e., $|G|/|F|$, where \( G = \{g \,|\, g \in F \setminus F' \text{ and } F' \notimplies g \} \))
    \item False Failure Rate (\Cref{def:ffr})
    \item Coverage on $E'$ (\Cref{def:coverage})
\end{enumerate}

Additionally, an important aspect of $\spade_{\text{sub}}$’s success is the effectiveness of subsumption assessment between all pairs of assertions. Since we do not have ground truth for subsumption, we focus on precision, calculated as the proportion of {\em correctly} identified subsumed pairs out of all subsumed pairs identified by the LLM. We do not assess recall---whether GPT-4 identified every possible subsumption---due to the impracticality of labeling possibly tens of thousands of assertion pairs per pipeline. Moreover, precision is more critical than recall or accuracy, as identifying even some subsumptions allows $\spade_{\text{sub}}$ to achieve a solution with fewer selected assertions than $\textsc{baseline}$.

\subsection{Results and Discussion}

\noindent
\begin{minipage}[t]{0.35\linewidth}
\footnotesize
    \centering
        \begin{tabular}{@{}lcc@{}}
            \toprule
            \textbf{Pipeline} & \textbf{Precision} \\ 
            \midrule
            codereviews       & 0.90 \\
            emails            & 0.79 \\
            fashion           & 0.74 \\
            finance           & 0.79 \\
            lecturesummaries  & 0.89 \\
            negotiation       & 0.68 \\
            sportroutine      & 0.89 \\
            statsbot          & 0.86 \\
            threads           & 0.80 \\
            \bottomrule
        \end{tabular}
       \captionof{table}{Precision of assessing subsumption with GPT-4 (verified by two authors).}
        \label{tab:subsumptionresults}
\end{minipage}\hfill
\begin{minipage}[t]{0.6\linewidth}
\centering
\footnotesize
        \begin{tabular}{@{}lcc@{}}
            \toprule
            \textbf{Pipeline} & {\bf $\spade_{\text{cov}}$} & {\bf $\spade_{\text{sub}}$} \\ 
            \midrule
            codereviews       & 0.267 & 0.362 \\
            emails            & 0.196 & 0.225 \\
            fashion           & 0.628 & 1.197 \\
            finance           & 0.352 & 0.441 \\
            lecturesummaries  & 0.265 & 0.538 \\
            negotiation       & 0.220 & 0.332 \\
            sportroutine      & 0.120 & 0.158 \\
            statsbot          & 0.104 & 0.122 \\
            threads           & 0.272 & 0.332 \\
            \bottomrule
        \end{tabular}
        \captionof{table}{Average ILP runtimes (in seconds) over 10 trials for $\spade_{\text{cov}}$ and $\spade_{\text{sub}}$. The $\textsc{baseline}$ method has no ILP component.}
        \label{tab:ilpruntimes}
\end{minipage}

First, we briefly discuss whether individual components of \spade are practical (e.g., determining subsumption, solving the ILPs). Using GPT-4 to assess subsumption results in an average precision of 0.82 across all pipelines, as seen in \Cref{tab:subsumptionresults}, confirming its effectiveness. The ILP runtimes in \Cref{tab:ilpruntimes} are all under one second, except for 1.197 seconds for $\spade_{\text{sub}}$ for the fashion pipeline. This relatively small runtime demonstrates the feasibility of our approach. In the remainder of this section, we will focus on the assertions chosen by different methods.

For simplicity, we set the coverage and FFR thresholds to be the same across all pipelines ($\alpha=0.6$, $\tau=0.25$). We report results for the three methods in \Cref{tab:spaderesults}. Consider the {\em codereviews} pipeline, for example, which uses an LLM to review a pull request for any code repository. Here, $\textsc{baseline}$ selects 20 assertions, $\spade_{\text{cov}}$ selects two assertions, and $\spade_{\text{sub}}$ selects 15 assertions. By selecting more functions, $\spade_{\text{sub}}$ ensures that all non-subsumed functions are included. All three approaches respect the $E'$ coverage constraint, 
but $\textsc{baseline}$ violates the FFR constraint in 4 out of 9 pipelines,
while the \spade approaches do not violate the FFR constraint. 

For our workloads, $\spade_{\text{sub}}$ opts for approximately 14\% fewer assertions compared to $\textsc{baseline}$ and shows a significantly lower FFR, reducing it by about 21\% relative to $\textsc{baseline}$. $\spade_{\text{cov}}$ excludes, on average, about 44\% of functions that are not subsumed by $F'$. Choosing a \spade implementation primarily depends on how much labeled data is available. We subsequently discuss the trade-offs between different \spade implementations.

\topic{Subsumption vs. $E'$ Coverage} $\spade_{\text{cov}}$ and $\spade_{\text{sub}}$ are complementary, the former being more useful if $E'$ is more comprehensive. For our datasets, $E'$ is not comprehensive: \Cref{tab:spaderesults} reveals that, on average, 44\% of functions excluded by $\spade_{\text{cov}}$ are not subsumed by the selected functions, despite being accurate within the FFR threshold. This is unsurprising given that each task has only 34 bad (i.e., failure) examples on average. While larger or more mature organizations may have extensive datasets and could get a meaningful result from $\spade_{\text{cov}}$, $\spade_{\text{sub}}$'s ability to select assertions that cover unrepresented {\em potential} failures can be beneficial in data-scarce settings. For example, here is a sample of 3 assertions for the {\em codereviews} pipeline ignored by $\spade_{\text{cov}}$ but included in $\spade_{\text{sub}}$ (with comments excluded for brevity):

\begin{mypython}
async def assert_includes_code_improvement_v2(
    example: dict, prompt: str, response: str
):
    question = "Does the response include suggestions for code improvements?"
    return await ask_llm(prompt, response, question)

async def assert_contains_brief_answers_v1(example: dict, prompt: str, response: str):
    question = "Is the response brief and to the point without unnecessary elaboration?"
    return await ask_llm(prompt, response, question)

async def assert_responds_to_correct_pull_request(
    example: dict, prompt: str, response: str
):
    pr_title = example["title"]
    question = (
        f"Is the response a review focused on the Pull Request titled '{pr_title}'?"
    )
    return await ask_llm(prompt, response, question)
\end{mypython}

\topic{Subsumption as a Means for Reducing Redundancy} Several pipelines exhibit a large discrepancy between functions selected in $\textsc{baseline}$ and $\spade_{\text{sub}}$, which occurs when there are many redundant candidates. For example, in the {\em codereviews} pipeline's 8 prompt versions, the developer iterated several times on the instruction to give a clear and concise review, resulting in five assertions that check the same thing (two of which are shown below):

\begin{mypython}
async def assert_response_is_concise_v1(
    example: dict, prompt: str, response: str
) -> bool:
    question = "Is the LLM response concise and to the point?"
    return await ask_llm(prompt, response, question)

async def assert_response_is_concise_and_clear(
    example: dict, prompt: str, response: str
):
    question = "Is this pull request review response concise and clear?"
    return await ask_llm(prompt, response, question)

async def assert_clear_professional_language_v1(
    example: dict, prompt: str, response: str
):
    question = "Is the response professional, clear, and without unnecessary jargon or overly complex vocabulary?"
    return await ask_llm(prompt, response, question)
\end{mypython}

Since all the five assertions meet the FFR constraint, individually, $\textsc{baseline}$ would select them all, which is undesirable because they all do the same thing, but $\spade_{\text{sub}}$ would select the one most compatible with the FFR constraint, as long as subsumption is assessed correctly. On the flip side, while assessing subsumption, the LLM may not recall all subsumptions, so $\spade_{\text{sub}}$ may have duplicate assertions. For example, the {\em codereviews} pipeline contains assertions titled \texttt{assert\_includes\_code\_improvement_v1} and \texttt{assert\_includes\_code\_improvement_v2}. 

\topic{$\boldsymbol{\alpha}$ and $\boldsymbol{\tau}$ Threshold Sensitivity} The feasibility of solutions from the ILP solver in \spade is dependent on the chosen $\alpha$ and $\tau$ thresholds. If a feasible solution is not found, developers may need to adjust these values in a binary search fashion. In our case, all 9 LLM pipelines yielded feasible solutions with $\alpha = 0.6$ and $\tau = 0.25$. However, the small size of $E'$ makes $\spade_{\text{cov}}$ particularly sensitive to $\alpha$. In the pipelines, we observed that between one and five assertions covered 60\% of $E'$'s failures. For example, $\spade_{\text{cov}}$ selected only one assertion for the {\em emails} pipeline:

\begin{mypython}
async def assert_encouragement_to_contact_company(
    example: dict, prompt: str, response: str
) -> bool:
    contact_phrases = [
        "reach out",
        "don't hesitate to contact",
        "looking forward to hearing from you",
        "if you have any questions",
        "need help getting started",
    ]
    return any(phrase in response for phrase in contact_phrases)
\end{mypython}

If $E'$ is exhaustive of failure modes and representative of the distribution of failures (e.g., for the {\em emails} pipeline, most failures are actually due to the response lacking an encouragement to contact the company), $\spade_{\text{cov}}$ might be a satisfactory solution. However, our $E'$ datasets clearly were not exhaustive, considering that $\spade_{\text{sub}}$ always chose additional assertions. $\spade_{\text{sub}}$ is less sensitive to $\alpha$, as it explicitly selects assertions based on their potential to cover new failures (i.e., subsumption) without exceeding the FFR, even if the constraint on coverage is no longer tight.

\topic{FFR Tradeoffs} Considering that the difference between the fraction of functions selected for $\textsc{baseline}$ and $\spade_{\text{sub}}$ is less than 10\% for three LLM pipelines, one may wonder if the complexity of $\spade_{\text{sub}}$ is worth it. $\spade_{\text{sub}}$ is generally preferable because $\textsc{baseline}$ fails to consistently meet the FFR threshold $\tau$. We observed that as prompt versions increase, so do the number of assertions, impacting $\textsc{baseline}$ adversely. The worst-case FFR of a set is the sum of individual FFRs, as shown in \Cref{eq:ffrbound}. Hence, with a large number of independent assertions, the total FFR is likely to surpass the threshold. This issue is evident in the {\em fashion} and {\em lecturesummaries} pipelines, where despite each of the 67 and 32 assertions meeting FFR constraints individually, the total FFR for $\textsc{baseline}$ reaches 88\% and 53\%, respectively. In practice, if \spade were to be deployed in an interactive system, where \spade could observe each LLM call in real-time (e.g., as a wrapper around the OpenAI API), the multitude of prompt versions further necessitates filtering assertions based on overall FFR. This underscores the need for the more complex $\spade_{\text{cov}}$ or $\spade_{\text{sub}}$ approaches.

\techreport{\subsection{Limitations and Future Work}\label{sec:experiments-limitations-future-work}

\topic{Improving Quality of LLMs} While LLMs (both closed and open-source) are improving rapidly, and we don't explicitly study prompt engineering strategies for \spade, a complementary research idea is to explore such strategies or fine-tune small open-source models to generate assertions. Moreover, we proposed subsumption as a coverage proxy but didn't explore prompt engineering or even non-LLM strategies (e.g., assertion provenance) for assessing subsumption. Despite LLM advancements, \spade's filtering stage remains crucial for reducing redundancy and ensuring accuracy, especially since assertions may involve LLMs. 

\topic{Collecting Labeled Examples} Acquiring labeled data ($E'$) is hard. Most of our datasets had very few prompt versions (only the ones committed to LangChain Hub), but in reality, developers may iterate on their prompt tens or hundreds of times. Future work could involve passive example collection via LLM API wrappers or gathering developer feedback on assertions. Prioritizing different types of assertions and formalizing these priorities within \spade is another area for exploration. Additionally, assessing the accuracy of FFR estimates with limited $E'$ and exploring methods to enhance FFR accuracy in the absence of large labeled datasets (e.g., via prediction-powered inference~\cite{angelopoulos2023prediction}), presents an interesting area for future work.

\topic{Supporting More Complex LLM Pipelines} Our study focuses on single-prompt LLM pipelines, but more complex pipelines, such as those involving multiple prompts, external resources, and human interaction, present opportunities for auto-generating assertions for each pipeline component. For instance, in retrieval-augmented generation pipelines\cite{lewis2020retrieval}, assertions could be applied to the retrieved context before LLM responses are even generated.}

%% file: figures/datasetdesc.tex
\begin{table*}[h]
\centering
\small
\setlength{\tabcolsep}{2pt}
\renewcommand{\arraystretch}{0.9}
\begin{threeparttable}
\begin{tabularx}{\textwidth}{@{}lcccXl@{}}
\toprule
\textbf{Pipeline} & \textbf{\# Good Ex.} & \textbf{\# Bad Ex.} & \textbf{\# Prompt Ver.} & \textbf{Data Generation Task} & \textbf{Full Prompt Link} \\ 
\midrule
\textit{codereviews} & 60 & 16 & 8  & Writing reviews of GitHub repo pull requests & \href{https://smith.langchain.com/hub/homanp/github-code-reviews}{homanp/github-code-reviews} \\
\textit{emails} &  43 & 55 & 3 & Creating SaaS user onboarding emails & \href{https://smith.langchain.com/hub/gitmaxd/onboard-email}{gitmaxd/onboard-email} \\
\textit{fashion} & 48 & 34 & 16 & Suggesting outfit ideas for specific events & N/A \\
\textit{finance}\tnote{a}  & 48 & 52 & 5 & Summarizing financial earnings call transcripts & \href{https://smith.langchain.com/hub/casazza/map_template}{casazza/map_template}  \\
\textit{lecturesummaries}\tnote{b}  & 27 & 22 & 14 &  Summarizing lectures or talks, focusing on main points and critical insights & \href{https://smith.langchain.com/hub/kirby/simple-lecture-summary}{kirby/simple-lecture-summary} \\
\textit{negotiation}  & 27 & 19 & 8 &  Writing tailored negotiation strategies based on provided contracts and target prices & \href{https://smith.langchain.com/hub/antoniogonc/strategy-report}{antoniogonc/strategy-report} \\
\textit{sportroutine}  & 19 & 31 & 3 & Transforming workout video transcripts into structured exercise routines & \href{https://smith.langchain.com/hub/aaalexlit/sport-routine-to-program}{aaalexlit/sport-routine-to-program} \\
\textit{statsbot} & 39 & 31 & 3 &  Writing interactive discussions for any topic in statistics & \href{https://smith.langchain.com/hub/anthonynolan/statistics-teacher}{anthonynolan/statistics-teacher} \\
\textit{threads} & 50 & 56 & 4 & Crafting concise, engaging Twitter threads for specific audiences and topics & \href{https://smith.langchain.com/hub/flflo/summarization}{flflo/summarization} \\
\bottomrule
\end{tabularx}
\begin{tablenotes}
    \item[a]{\url{https://www.kaggle.com/datasets/ashwinm500/earnings-call-transcripts}} \quad \textsuperscript{b} {\url{https://www.kaggle.com/datasets/miguelcorraljr/ted-ultimate-dataset}}
\end{tablenotes}
\end{threeparttable}
\caption{Description of data-generating LLM pipelines in our experiments. The fashion examples (and ground-truth indicators of whether the example response is good or bad) are provided by a startup that uses LangChain. All other examples are synthetically generated except examples for the {\em finance} and {\em lecturesummaries} pipelines, which are taken from Kaggle.}
\label{tab:dataset-desc}
\end{table*}

%% file: figures/spaderesults.tex
\begin{table}
\centering
\footnotesize
\setlength{\tabcolsep}{4pt} 
\begin{tabularx}{\linewidth}{@{}>{\raggedright\arraybackslash}rccc@{\hspace{2pt}}lc@{\hspace{2pt}}l@{\hspace{2pt}}r@{\hspace{2pt}}lr@{\hspace{2pt}}l@{}}
\toprule
\textbf{Pipeline} & \textbf{\# CA} &  \textbf{Method} & \multicolumn{2}{c}{\textbf{FFR}}  & \multicolumn{2}{c}{\makecell{\textbf{Coverage}\\\textbf{on $E'$}}}   & \multicolumn{2}{c}{\makecell{\textbf{Frac Func.} \\ \textbf{Selected}}} & \multicolumn{2}{c}{\makecell{\textbf{Frac Excl.} \\ \textbf{ Funcs. not} \\ \textbf{Subsumed}}} \\ 
\midrule
\multirow{3}{*}{{ \em codereviews}} & \multirow{3}{*}{44} & $\textsc{baseline}$ & 0.117 & \cmark & 1 & \cmark & 0.456 & (20) & 0 & (0) \\
& & $\spade_{\text{cov}}$ & 0 & \cmark & 0.625 & \cmark & 0.045 & (2) & 0.409 & (18) \\
& & $\spade_{\text{sub}}$ & 0.117 & \cmark & 0.875 & \cmark & 0.341 & (15) & 0 & (0) \\
\midrule
\multirow{3}{*}{{\em emails}} & \multirow{3}{*}{24} & $\textsc{baseline}$ & 0 & \cmark & 1 & \cmark & 0.5 & (12) & 0 & (0) \\
& & $\spade_{\text{cov}}$ & 0 & \cmark & 1 & \cmark & 0.0417 & (1) & 0.458 & (11) \\
& & $\spade_{\text{sub}}$ & 0 & \cmark & 1 & \cmark & 0.458 & (11) & 0 & (0) \\
\midrule
\multirow{3}{*}{{\em fashion}} & \multirow{3}{*}{106} & $\textsc{baseline}$ & 0.878 & \xmark & 0.971 & \cmark & 0.632 & (67) & 0 & (0) \\
& & $\spade_{\text{cov}}$ & 0.245 & \cmark & 0.6 & \cmark & 0.028 & (3) & 0.321 & (34) \\
& & $\spade_{\text{sub}}$ & 0.224 & \cmark & 0.62 & \cmark & 0.377 & (40) & 0 & (0) \\
\midrule
\multirow{3}{*}{{\em finance}} & \multirow{3}{*}{47} & $\textsc{baseline}$ & 0.667 & \xmark & 1 & \cmark & 0.787 & (37) & 0 & (0) \\
& & $\spade_{\text{cov}}$ & 0.229 & \cmark & 0.673 & \cmark & 0.085 & (4) & 0.553 & (26) \\
& & $\spade_{\text{sub}}$ & 0.208 & \cmark & 0.981 & \cmark & 0.553 & (26) & 0 & (0) \\
\midrule
\multirow{3}{*}{{\em lecturesum.}} & \multirow{3}{*}{70} & $\textsc{baseline}$ & 0.528 & \xmark & 1 & \cmark & 0.457 & (32) & 0 & (0) \\
& & $\spade_{\text{cov}}$ & 0.194 & \cmark & 0.643 & \cmark & 0.014 & (1) & 0.414 & (29) \\
& & $\spade_{\text{sub}}$ & 0.194 & \cmark & 1 & \cmark & 0.343 & (24) & 0 & (0) \\
\midrule
\multirow{3}{*}{{\em negotiation}} & \multirow{3}{*}{50} & $\textsc{baseline}$ & 0.444 & \xmark & 1 & \cmark & 0.4 & (20) & 0 & (0) \\
& & $\spade_{\text{cov}}$ & 0.222 & \cmark & 0.632 & \cmark & 0.04 & (2) & 0.32 & (16) \\
& & $\spade_{\text{sub}}$ & 0.185 & \cmark & 1 & \cmark & 0.34 & (17) & 0 & (0) \\
\midrule
\multirow{3}{*}{{\em sportroutine}} & \multirow{3}{*}{26} & $\textsc{baseline}$ & 0.211 & \cmark & 1 & \cmark & 0.538 & (14) & 0 & (0) \\
& & $\spade_{\text{cov}}$ & 0.211 & \cmark & 0.774 & \cmark & 0.077 & (2) & 0.462 & (12) \\
& & $\spade_{\text{sub}}$ & 0 & \cmark & 0.871 & \cmark & 0.308 & (8) & 0 & (0) \\
\midrule
\multirow{3}{*}{{\em statsbot}} & \multirow{3}{*}{15} & $\textsc{baseline}$ & 0 & \cmark & 1 & \cmark & 0.467 & (7) & 0 & (0) \\
& & $\spade_{\text{cov}}$ & 0 & \cmark & 0.935 & \cmark & 0.133 & (2) & 0.333 & (5) \\
& & $\spade_{\text{sub}}$ & 0 & \cmark & 1 & \cmark & 0.467 & (7) & 0 & (0) \\
\midrule
\multirow{3}{*}{{\em threads}} & \multirow{3}{*}{34} & $\textsc{baseline}$ & 0 & \cmark & 1 & \cmark & 0.765 & (26) & 0 & (0) \\
& & $\spade_{\text{cov}}$ & 0 & \cmark & 0.875 & \cmark & 0.029 & (1) & 0.735 & (25) \\
& & $\spade_{\text{sub}}$ & 0 & \cmark & 1 & \cmark & 0.589 & (20) & 0 & (0) \\
\bottomrule
\end{tabularx}
\caption{Results of different versions of \spade with $\alpha=0.6$ and $\tau=0.25$. ``\# CA'' is short for the number of candidate assertions. The \cmark and \xmark~marks denote whether $\alpha$ and $\tau$ constraints are met. Each entry is a fraction of the total number of candidate assertions for that pipeline (with the absolute number in parentheses). $\spade_{\text{cov}}$ selects the fewest assertions overall. $\spade_{\text{sub}}$ selects the fewest assertions while optimizing for subsumption.}
\label{tab:spaderesults}
\end{table}

%% file: sections/related.tex

We survey work from prompt engineering, evaluating ML and LLMs, LLMs for software testing, and testing ML pipelines.


\topic{Prompt Engineering} For both nontechnical~\cite{zamfirescu2023johnny} and technical users~\cite{si2022prompting, parameswaran2023revisiting}, prompt engineering is hard for several reasons: small changes in prompt phrasing~\cite{liu2023pre, arora2022ask} or the order of instructions or contexts~\cite{lu2021fantastically} can significantly affect outputs. Moreover, as LLMs change under the hood of the API (i.e., prompt drift), outputs can change without developer awareness~\cite{chen2023chatgpt}. Tools and papers are emerging to aid in prompt management and experimentation, and are even using LLMs to write prompts~\cite{arawjo2023chainforge, wu2022ai, wu2022promptchainer, cheng2023prompt, zhou2022large}. Moreover, {\em deployed} prompts introduce new challenges, like ``balancing more context with fewer tokens'' and ``wrangling prompt output'' to meet user-defined criteria~\cite{parnin2023building}. Our work doesn't focus explicitly on helping developers create better prompts, but it could indirectly support developers by helping identify mistakes.

\topic{ML and LLM Evaluation} Evaluating and monitoring deployed ML models is known to be challenging~\cite{paleyes2022challenges, shankar2022rethinking}. Evaluating LLMs in a deployed setting is even more challenging because LLMs are typically used for generative tasks, where the outputs are free-form~\cite{dou2021gpt}. Some LLM pipeline types, like question-answering with retrieval-augmented generation pipelines~\cite{lewis2020retrieval}, can use standardized, automated metrics~\cite{es2023ragas, saad2023ares},  but others face challenges due to unknown metrics and lack of labeled datasets~\cite{xiao2023evaluating, chang2023survey, parnin2023building}. Typically, organizations rely on human evaluators for LLM outputs~\cite{gehrmann2023repairing, wang2023aligning, parnin2023building}, but recent studies suggest LLMs can self-evaluate effectively with detailed ``scorecards''~\cite{chan2023chateval, zhang2023wider, kim2023evallm}. However, writing these scorecards can be challenging~\cite{parnin2023building}, motivating auto-generated evaluators. Recent work~\cite{rebedea2023nemo, singhvi2023dspy, kim2023evallm} and industry tools~\cite{guardrails,langchain,llama-index} proposes the use of assertions
to catch mistakes in LLM pipelines,
while requiring the user to select these assertions.

\topic{LLMs for Software Testing} LLMs are increasingly being used in software testing, mainly for generating unit tests and test cases~\cite{wang2023software, schafer2023empirical, siddiq2023exploring, steenhoek2023reinforcement, lemieux2023codamosa}. Research explores how LLMs' prompting strategies, hallucinations, and nondeterminism affect code or test accuracy~\cite{ouyang2023llm, feldt2023towards, chen2021evaluating, fan2023large}. Our work is complementary and leverages LLMs to generate code-based assertions for LLM pipelines.

\topic{Testing and Validation in ML Pipelines} ML pipelines are hard to manage in production. Much of the literature on ML testing is geared towards validating structured data, through analyzing data quality~\cite{datalinter, breckdataval, schelterdataval, shankar2023automatic} or provenance~\cite{arguseyes, vamsa}. Platforms for ML testing typically offer automated experiment tracking and prevention against overfitting~\cite{easeml,mldp}, as well as tools for data distribution debugging~\cite{grafberger2022data}. Model-specific assertions typically require human specification~\cite{kang2020model}, or at least large amounts of data to train learned assertions~\cite{kang2022finding}. LLM chains or pipelines are a new class of ML pipelines, and LLMs themselves can generate assertions with little data. A recent study highlights the difficulty of testing LLM pipelines for "copilot"-like products: developers want to ensure accuracy while avoiding excessive resource use, such as running hundreds of assertions~\cite{parnin2023building}---motivating assertion filtering.

%% file: sections/app.tex
\appendix
\input{figures/chains}

\section{Notation}
\label{app:notation}

\Cref{tab:notation} summarizes the important notation used in the paper, including high-level descriptions of the ILP variables.

\begin{table}[H]
\centering
\footnotesize
\begin{tabularx}{\linewidth}{c|X}
\toprule
\textbf{Symbol} & \textbf{Description} \\ 
\midrule
$\mathcal{P}$ & A prompt template (i.e., a string with placeholders, intended to be submitted to an LLM) \\ 
$\mathcal{P}_i$ & The \(i\)th version of a prompt template \\ 
$\Delta \mathcal{P}_i$ & The diff or changes between consecutive versions of a prompt template \\ 
$E$ & The set of all hypothetical example runs for an LLM pipeline \\ 
$E'$ & A small set of example runs for an LLM pipeline (with labels denoting whether responses were good) that we can observe, where $E' \subset E$ \\ 
$n$ & The number of examples in $E'$ \\
$e_i$ & The \(i\)th example in $E'$ \\ 
$f$ & An arbitrary assertion function \\ 
$F$ & The set of all candidate assertions from \Cref{sec:candidateassertions} \\ 
$m$ & The number of assertions in $F$ \\
$F'$ & A subset of $F$ selected as a minimal set of assertions \\ 
$G$ & Set of functions in $F \setminus F'$ not subsumed by $F'$ (as defined by \Cref{def:subsumption}) \\ 
$\hat{y}_i$ & Binary indicator of whether example $e_i$ satisfies all assertions in $F'$ \\ 
$y_i$ & Binary indicator of whether a developer considers the LLM pipeline response for example $e_i$ good (i.e., success) or bad (i.e., failure) \\ 
$\alpha$ & Threshold for coverage of failing examples; $F'$ should cover at least $\alpha$ of failing examples \\ 
$\tau$ & Threshold for False Failure Rate (FFR); $F'$ should fail {\em no more than} $\tau$ good examples  \\ 
$M$ & $n \times m$ matrix representing the results of assertions on examples in $E'$ \\ 
$K$ & $m \times m$ matrix representing subsumption relationships between functions \\ 
$x_j$ & ILP variable indicating inclusion of $f_j$ in $F'$ \\ 
$w_{ij}$ & ILP variable representing if $F'$ denotes $e_i$ as a failure \\ 
$u_i$ & ILP variable indicating if a failed example is covered \\ 
$z_i$ & ILP variable defining if $F'$ incorrectly marks a successful example as a failure (false failure) \\
$r_j$ & ILP variable indicating if a function is subsumed by any function in $F'$ \\
$s_j$ & ILP variable representing functions in $F \setminus F'$ not subsumed by $F'$ \\
\bottomrule
\end{tabularx}
\caption{Notation used in the paper}
\label{tab:notation}
\end{table}

\section{LangChain Prompt Hub and Taxonomy of Prompt Deltas}
\label{app:chains}

The LangChain Prompt Hub is an open-source repository of prompts for chains, detailing the version history of the prompts. In analyzing prompt deltas, we filter the Prompt Hub for user-uploaded chains with at least 3 prompt versions. \Cref{tab:chains} summarizes the chains analyzed.

\section{\spade Prompts}
\label{app:prompts}

\spade leverages LLMs in three places. The first uses GPT-4 to categorize the delta (constructed by Python's \texttt{difflib}). The second uses GPT-4 to generate Python assertion functions, based on the prompt and categories identified by the first ste.. The third usage of LLMs is in asking GPT-4 to evaluate whether two functions subsume each other.

Given \texttt{prompt\_diff}, a list of sentences that have been modified from the previous prompt template to the current prompt template (i.e., $\Delta\mathcal{P}$), the prompt for categorizing the delta is as follows:

\begin{mypython}
"""
Here are the changed lines in my prompt template:

``{prompt_diff}''

I want to write assertions for my LLM pipeline to run on all pipeline responses. Here are some categories of assertion concepts I want to check for:

- Presentation Format: Is there a specific format for the response, like a comma-separated list or a JSON object?
- Example Demonstration: Does the prompt template include any examples of good responses that demonstrate any specific headers, keys, or structures?
- Workflow Description: Does the prompt template include any descriptions of the workflow that the LLM should follow, indicating possible assertion concepts?
- Count: Are there any instructions regarding the number of items of a certain type in the response, such as ``at least'', ``at most'', or an exact number?
- Inclusion: Are there keywords that every LLM response should include?
- Exclusion: Are there keywords that every LLM response should never mention?
- Qualitative Assessment: Are there qualitative criteria for assessing good responses, including specific requirements for length, tone, or style?
- Other: Based on the prompt template, are there any other concepts to check in assertions that are not covered by the above categories?

Give me a list of concepts to check for in LLM responses. Each item in the list should contain a string description of a concept to check for, its corresponding category, and the source, or phrase in the prompt template that triggered the concept. For example, if the prompt template is "I am a still-life artist. Give me a bulleted list of colors that I can use to paint <object>.", then a concept might be "The response should include a bulleted list of colors." with category "Presentation Format" and source "Give me a bulleted list of colors".

Your answer should be a JSON list of objects within ```json ``` markers, where each object has the following fields: "concept", "category", and "source". This list should contain as many assertion concepts as you can think of, as long are specific and reasonable.
"""
\end{mypython}

Let \texttt{concepts} be the parsed categories identified by the previous prompt. The prompt for generating the Python assertion functions is as follows:

\begin{mypython}
"""
Here is my prompt template:

"{prompt_template}"

Here is an example and its corresponding LLM response:

Example: {sample_example}
LLM Response: {sample_response}

Here are the concepts I want to check for in LLM responses:

{concepts}

Give me a list of assertions as Python functions that can be used to check for these concepts in LLM responses. Assertion functions should not be decomposed into helper functions. Assertion functions can leverage the external function `ask_llm` if the concept is too hard to evaluate with Python code alone (e.g., qualitative criteria). The `ask_llm` function accepts formatted_prompt, response, and question arguments and submits this context to an expert LLM, which returns True or False based on the context. Since `ask_llm` calls can be expensive, you can batch similar concepts that require LLMs to evaluate into a single assertion function, but do not cover more than two concepts with a function. For concepts that are ambiguous to evaluate, you should write multiple different assertion functions (e.g., different `ask_llm` prompts) for the same concept(s).

Each function should take in 3 args: an example (dict with string keys), prompt formatted on that example (string), and LLM response (string). Each function shold return a boolean indicating whether the response satisfies the concept(s) covered by the function. Here is a sample assertion function for an LLM pipeline that generates summaries:

```python
def assert_simple_and_coherent_narrative(example: dict, prompt: str, response: str):
    # Check that the summary form a simple, coherent narrative telling a complete story.

    question = "Does the summary form a simple, coherent narrative telling a complete story?"
    return ask_llm(prompt, response, question)
```

Your assertion functions should be distinctly and descriptively named, and they should include a docstring describing what the function is checking for.
"""
\end{mypython}






\newpage
\subsection{Evaluating Subsumption}

To evaluate subsumption, we use one prompt to query the subsumed pairs, given all assertions, and then a second prompt to format the subsumed pairs as a JSON so it can be easily parsed by \spade. The first prompt is as follows:

\begin{mypython}
"""
Here are all the functions I have:\n\n{assertion_blob}\n\nBased on the code, please identify every pair of functions where one function implies the other. Note that function A might imply function B, but function B may not imply function A. If two functions A and B check for the same thing, then they both imply each other (i.e., A implies B and B implies A), so you should list both directions. Feel free to use the function names to decide if two functions check for the same thing.
"""
\end{mypython}

In the prompt above, the \texttt{assertion\_blob} represents a string of all assertion functions. Then, the second prompt is as follows:

\begin{mypython}
"""
Please return your answer as a JSON list within ```json ``` ticks, where each element of the list is a tuple (A, B). If two functions A and B check for the same thing, make sure to include both tuples (A, B) and (B, A). For example, if I only had two functions `check_json` and `assert_json`, the answer should be: ```json\n[("check_json", "assert_json"), ("assert_json", "check_json")]```
"""
\end{mypython}

%% file: figures/chains.tex
\begin{table*}[]
    \centering
    \footnotesize
    \begin{tabular}{l|l|c}
\toprule
{\bf Task Domain} & {\bf Summary of Prompt} & {\bf Num. Versions}\\
\midrule
Conversational AI & Act as an AI assistant that can execute tools and have a natural conversation with a user. & 8\\
Web Development & Generate Tailwind CSS components like text, tables, and cards to help answer fantasy football questions. & 3\\
Question Answering & Identify key assumptions in questions and generate follow-up questions to fact check those assumptions. & 5\\
Programming Assistant & Safely execute any code a user provides to help them complete tasks, while alerting them to any concerning instructions. & 3\\
Model Evaluation & Evaluate a model's outputs by assigning a score based on provided criteria and examples. & 6\\
Question Answering & Concisely answer questions using no more than 3 sentences and provided context passages. & 9\\
Workflow Automation & Create a JSON workflow using a list of provided tools based on the user's natural language query. & 11\\

Question Answering & Answer open-ended questions by asking clarifying follow-up questions before providing a final answer. & 8\\
Information Retrieval & Determine if a passage contains enough useful information to help answer a specific question. & 3\\
Code Translation & Convert Python code snippets to valid, idiomatic TypeScript code. & 6\\
Code Review & Review GitHub pull requests and provide constructive feedback for improvement. & 8\\
Question Answering & Concisely answer questions using no more than 3 sentences and provided context passages. & 5\\
Email Marketing & Craft a user onboarding email following marketing best practices based on provided context. & 3\\
Text Summarization & Summarize long text into a compelling, engaging Twitter thread for a target audience. & 4\\
Email Marketing & Craft a user onboarding email following marketing best practices based on provided context. & 7\\
Procurement Automation & Develop a detailed, tailored negotiation strategy report using provided information about suppliers, goals, etc. & 8\\
Education & Teach statistics topics interactively by answering questions, providing feedback, and posing example problems. & 3\\
Fitness & Convert a text description of a fitness challenge into a structured exercise program. & 3\\
Education & Generate engaging, concise quiz questions based on the information contained in a provided context document. & 5\\
\bottomrule
\end{tabular}
    \caption{Description of each chain in our dataset. We describe the domain of the task the chain is trying to perform, and a short summary of the task.}
    \label{tab:chains}
\end{table*}